\documentclass[draftcls,onecolumn,letterpaper,twoside]{IEEEtran}

\usepackage[T1]{fontenc}
\usepackage[latin9]{inputenc}
\usepackage{verbatim}
\usepackage{amsmath}
\usepackage{graphicx}
\usepackage{cite}
\usepackage{amssymb}
\usepackage{esint}
\usepackage{ae}

\makeatletter

\newcommand{\lyxdot}{.}

  \newtheorem{lemma}{Lemma}
  \newtheorem{thm}{Theorem}
  \newtheorem{prop}{Proposition}

\usepackage[english]{babel}

\makeatother
\begin{document}
 
\def \var{\operatorname{var}} 
\def \cov{\operatorname{cov}}
\newcommand{\eg}{\textit{e.g.}} 
\newcommand{\ie}{\textit{i.e.}}
\newcommand{\N}{N}
\newcommand{\M}{M}
\newcommand{\K}{K} 
\def\area{\operatorname{area}}
\def\figref#1{Fig.\,\ref{#1}}%
\def\E{\mathbb{E}} 
\def\d{\text{d}}
\def\V{\mathbb{V}} 
\def\vv{\varrho^{(2)}} 
\def\P{\mathbb{P}} 
\def\R{\mathbb{R}}
\def\Z{\mathbb{Z}}
\def\X{\mathcal{X}} 
\def\RP{\mathbb{R}\times\mathbb{P}}
 \def\poi{\text{poi}}

\title{Interference and Outage in Clustered Wireless Ad Hoc Networks }

\author{Radha Krishna Ganti and Martin Haenggi\\
Department of Electrical Engineering\\
University of Notre Dame\\
Indiana-46556, USA\\
E-mail \texttt{\{rganti,mhaenggi\}@nd.edu}
\thanks{Part of the material in this paper has been presented at the 2006
Asilomar conference.}} 

\maketitle
 
\begin{abstract}
In the analysis of large random  wireless
networks, the underlying node distribution is almost ubiquitously
assumed to be the homogeneous Poisson point process. In this paper,
the node locations are assumed to form a \emph{Poisson clustered process}
on the plane. We derive the distributional properties of the interference
and provide upper and lower bounds for its CCDF.  
  We consider the probability of successful transmission   in an interference limited channel when
fading is modeled as Rayleigh. We provide a numerically integrable
expression for the outage probability and closed-form upper and lower
bounds. 
We show that when the transmitter-receiver distance is large, the
success probability is greater than that of a Poisson arrangement.
These results characterize the performance of the system under geographical
or MAC-induced clustering. We obtain the maximum intensity of transmitting
nodes for a given outage constraint, \ie, the transmission capacity
(of this spatial arrangement) and show that it is equal to that of
a Poisson arrangement of nodes. For the analysis, techniques from
stochastic geometry are used, in particular the probability generating
functional of Poisson cluster processes, the Palm characterization
of Poisson cluster processes and the Campbell-Mecke theorem. 
\end{abstract}

\section{Introduction}

A common and analytically convenient assumption for the node distribution
in large wireless networks is the homogeneous (or stationary) Poisson
point process (PPP) of intensity $\lambda$, where the number of nodes
in a certain area of size $A$ is Poisson with parameter $\lambda A$,
and the numbers of nodes in two disjoint areas are independent random
variables. For sensor networks, this assumption is usually justified
by claiming that sensor nodes may be dropped from aircraft in large
numbers; for mobile ad hoc networks, it may be argued that terminals
move independently from each other. While this may be the case for
certain networks, it is much more likely that the node distribution
is not \char`\"{}completely spatially random\char`\"{} (CSR), \ie,
that nodes are either clustered or more regularly distributed. Moreover,
even if the complete set of nodes constitutes a PPP, the subset of
{\em active} nodes (\eg, transmitters in a given time-slot or sentries
in a sensor network), may not be homogeneously Poisson. Certainly,
it is preferable that simultaneous transmitters in an ad hoc network
or sentries in a sensor network form more regular processes to maximize
spatial reuse or coverage respectively. On the other hand, many protocols
have been suggested that are based on clustered processes. This motivates
the need to extend the rich set of results available for PPPs to other
node distributions. The clustering of nodes may be due to geographical
factors, for example communicating nodes inside a building or groups
of nodes moving in a coordinated fashion. The clustering may also
be {}``artificially'' induced by MAC protocols. We   denote
the former as geographical clustering and the latter as logical clustering.

\subsection{Related Work }

\label{sec:Related-Work}

There exists a significant body of literature for networks with Poisson
distributed nodes. In \cite{Sousa90} the characteristic function
of the interference was obtained when there is no fading and the nodes
are Poisson distributed. They also provide the probability distribution
function of the interference as an infinite series. Mathar et al.,
in \cite{mathar1995dci}, analyze the interference when the interference
contribution by a transmitter located at $x$, to a receiver located
at the origin is exponentially distributed with parameter $\Vert x\Vert^{2}$.
Using this model they derive the density function of the interference 
when the nodes are arranged as a one dimensional lattice. Also the
Laplace transform of the interference is obtained when the nodes are
Poisson distributed. 

It is known that the interference in a planar network of nodes can be
modeled as a shot noise process. Let $\{x_{j}\}$ be a point process
in $\mathbb{R^{d}}$. Let $\{\beta_{j}(.)\}$ be a sequence of independent
and identically distributed random functions on $\mathbb{R}^{d}$,
independent of $\{x_{j}\}$. Then a generalized shot noise process
can be defined as \cite{westcott}\[
Y(x)=\sum_{j}\beta_{j}(x-x_{j})\]
If  $\beta_{j}()$ is the path loss model with fading, $Y(x)$ is
the interference at location $x$ if all nodes $x_{j}$ are transmitting.
The shot noise process is a very well studied process for noise modeling.
It was first introduced by Schottky in the study of fluctuations in
the anode current of a thermionic diode and it was studied in detail
by Rice \cite{rice1954mar,rice1977gsn}. Daley in $1971$ defined
multi-dimensional shot noise and examined its existence when the points
$\{x_{j}\}$ are Poisson distributed in $\mathbb{R}^{d}$. The existence
of generalized shot-noise process, for any point process was studied
by Westcott in \cite{westcott}. Westcott also provides the Laplace
transform of the shot-noise when the points $\{x_{j}\}$ are distributed
as a Poisson cluster process. Normal convergence of the multidimensional
shot-noise process is shown by Heinrich and Schmidt \cite{heinrich1985ncm}.
They also show that when the points $\{x_{j}\}$ form a Poisson point
process of intensity $\lambda$, the rate of convergence to a normal
distribution is $\sqrt{\lambda}$.

In \cite{ilow1998aas}, Ilow and Hatzinakos model the interference
as a shot noise process and show that the interference is a symmetric
$\alpha$-stable process \cite{shao1993spf} when the nodes are Poisson
distributed on the plane. They also show that channel randomness affects
the dispersion of the distribution, while the path-loss exponent affects
the exponent of the process. The throughput and outage in the presence
of interference are analyzed in \cite{bacelli-aloha,jagadish04allerton,jagadish06allerton}.
In \cite{bacelli-aloha}, the shot-noise process is analyzed using
stochastic geometry when the nodes are distributed as Poisson and
the fading is Rayleigh. In \cite{WebAnd2006a} upper and lower bounds
are obtained under general fading and Poisson arrangement of nodes. 

Even in the case of the PPP, the interference distribution is not
known for all fading distributions and all channel attenuation models.
Only the characteristic function or the Laplace transform of the interference
can be obtained in most of the cases. The Laplace transform can be
used to evaluate the outage probabilities under Rayleigh fading characteristics
\cite{bacelli-aloha,haenggi2005isit}. In the analysis of outage probability,
the \emph{conditional} Laplace transform is required, \ie, the Laplace
transform given that there is a point of the process located at the origin.
For the PPP, the conditional Laplace transform is equal to the unconditional Laplace
transform. To the best of our knowledge, we are not aware of any literature
 pertaining to the interference characterization in a clustered network.

\cite{weber:2005} introduces the notion of {\em transmission capacity},
which is a measure of the area spectral efficiency of the successful
transmissions resulting from the optimal contention density as a function
of the link distance. 
Transmission capacity is defined as the product of the maximum density
of successful transmissions and their data rate, given an outage constraint.
Weber et al., provide bounds for the transmission capacity under different
models of fading, when the node location are Poisson distributed.

\subsection{Main contributions and organization of the paper}

\begin{figure}
\begin{centering}
\includegraphics[scale=0.5]{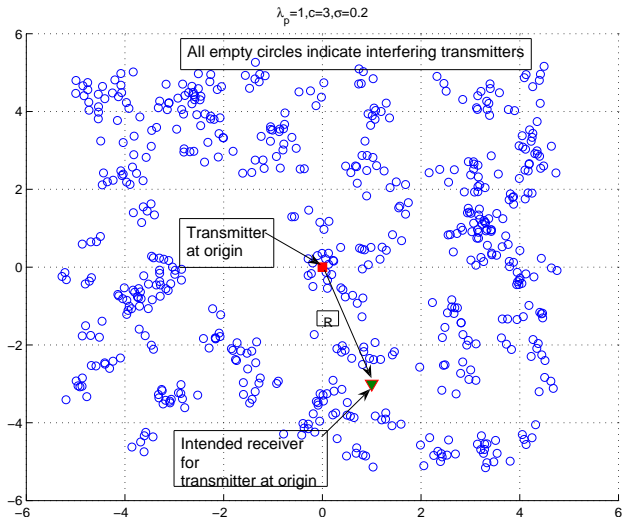}
\par\end{centering}

\begin{centering}
\caption{Illustration of transmitters and receivers. Cluster density is $1$.  Transmitter density in each cluster  is $3$. Spread of each cluster  is Gaussian with standard deviation $\sigma=0.25$. Observe that the intended receiver for the transmitter at the origin is not a part of the cluster process. The transmitter at the origin is a part of the cluster located around the origin.}
\label{fig:fig1}
\par\end{centering}

\end{figure}

In this work, we model the transmitters as a Poisson cluster process.
To circumvent technical difficulties we assume that the receivers
are not a part of this clustered process. We then focus on a specific
transmit-receive pair at a distance $R$ apart, see Fig \ref{fig:fig1}.
We evaluate the Laplace transform of the interference on the plane
conditioned on the event that there is a transmitter located at the
origin. Upper and lower bounds are obtained for the CCDF of the interference.
From these bounds, it is observed that the interference is a heavy-tailed
distribution with exponent $2/\alpha$ when the path loss function
is $\Vert x\Vert^{-\alpha}$. When the path-loss function has no singularity at the origin (\ie, remains bounded),
 the distribution of interference depends heavily on the fading distribution.
 Using the Laplace transform, the probability
of successful transmission between a transmitter and receiver in an
interference-limited  Rayleigh channel is obtained.
We provide a numerically integrable expression for the outage probability
and closed-form upper and lower bounds. The {\em clustering gain} $G(R)$
is defined as the ratio of success probabilities  of  the  clustered process  and the PPP  with the same intensity. It
is observed that when the transmitter-receiver distance $R$ is large,
the clustering gain $G(R)$ is greater than unity and becomes infinity
as $R\rightarrow\infty$. The gain $G(R)$ at small $R$ depends on
the path loss model and the total intensity of transmissions. We provide
conditions on the total intensity of transmitters under which the
gain is greater than unity for small $R$. This is useful to determine
when logical clustering performs better than uniform deployment of
nodes. We also obtain the maximum intensity of transmitting nodes
for a given outage constraint, \ie, the transmission capacity \cite{weber:2005,WebAnd2006a,WebAnd2006d}
of this spatial arrangement and show that it is equal to that of a
Poisson arrangement of nodes.   We observe that in a spread-spectrum
system, clustering is beneficial for long range transmissions, and we compare DS-CDMA and FH-CDMA.%
{}

The paper is organized as follows: in Section \ref{sec:System-Model}
we present the system model and assumptions, introduce the Neyman-Scott
cluster process and derive its conditional generating functional.
In Section \ref{sec:Outage-Probability} we derive the properties
of interference, outage probability and the gain function $G(R)$.
In Section \ref{sec:Transmission-Capacity}, we derive the transmission
capacity of the clustered network. %
{}

\section{System Model and Assumptions\label{sec:System-Model}}

In this section we introduce the system model and derive some required
results for the Poisson cluster process.

\subsection{System model and notation}

The location of transmitting nodes is modeled as a stationary and
isotropic Poisson cluster process $\phi $ on $\mathbb{R}^{2}$.
The receiver is not considered a part of the process. See Figure \ref{fig:fig1}.
Each transmitter   is assumed to transmit at unit
power. The power received by a receiver located at $z$ due to a transmitter
at $x $ is modeled as $h_xg(x -z)$, where $h_x$ is
the power fading coefficient (square of the amplitude fading coefficient) associated with the channel between the nodes
$x $ and $z$. We also assume that all the fading coefficients
are independent and are drawn from the same distribution.  We
will sometimes use $h$ to denote a random variable that is i.i.d
with the power fading coefficients.   Let $\{o\}$ denote the origin $(0,0)$. We assume that the path loss model
$g(x):\ \R^{2}\setminus\{o\}\rightarrow\R^{+}$ satisfies the following
conditions.

\begin{enumerate}
\item $g(x)$ is a continuous, positive, non-increasing function of $\Vert x\Vert$   and  \[
\int_{\R^{2}\setminus B(o,\epsilon)}g(x)\d x<\infty,\quad \forall\epsilon>0\]
 where $B(o,\epsilon)$ denotes a ball of radius $\epsilon$ around
the origin.
\item \begin{equation}
\lim_{\Vert x\Vert\rightarrow\infty}\frac{g(x)}{g(x-y)}=1,\quad  \forall y\in\R^{2}\label{eq:assume2}\end{equation}
 
\end{enumerate}
 
$g(x)$ is usually taken to be a power law in the form $\Vert x\Vert^{-\alpha}$,
$(1+\Vert x\Vert^{\alpha})^{-1}$ or $\min\{1,\Vert x\Vert^{-\alpha}\}$.
To satisfy condition $1$, we require $\alpha>2$. The interference
at node $z$ on the plane is given by \begin{eqnarray}
I_{\phi}(z) & = & \sum_{x \in\phi}h_xg(x -z)\label{eq:interfernce-main}\end{eqnarray}
 The conditions required for the existence of $I_{\phi}(z)$ are discussed
in~\cite{westcott}. Let $W$ denote the additive Gaussian noise
the receiver. We say that the communication from a transmitter at the 
origin to a receiver situated at $z$ is successful if and only if
\begin{eqnarray}
\frac{h g(z)}{W+I_{\phi\setminus\{x\}}(z)} & \geq & T\end{eqnarray}
 or equivalently, \[
\frac{h g(z)}{W+I_{\phi}(z)}\geq\frac{T}{1+T}\]
For the calculation of outage probability and transmission capacity,
 the amplitude fading $\sqrt{h_{x}}$ is assumed to be Rayleigh with mean $\mu$, but some results
are presented for the more general case of Nakagami-$m$ fading. Hence
the powers $h_{x}$ are exponentially and gamma distributed respectively.
We will be evaluating the performance of spread-spectrum in some sections
of the paper. Even though we evaluate spread-spectrum systems (specifically
DS-CDMA and FH-CDMA) we will not be using any power control, the reason
being that there is no central base station.

 Notation:   If $\lim_{x\rightarrow\infty}f(x)/g(x)=C$, we shall use $f(x)\sim g(x)$ if $C=1$, $f(x) \lesssim g(x)$ if $0<C<1$ and $f(x) \gtrsim g(x)$  if $1<C<\infty$.

\subsection{\label{sec:Neyman-Scott-cluster-process}Neyman-Scott cluster processes}

Neyman-Scott cluster processes~\cite{stoyan} are Poisson cluster
processes that result from homogeneous independent clustering applied
to a stationary Poisson process, where the parent points form a stationary
Poisson process $\phi_{p}=\{x_{1},x_{2},\ldots\}$ of intensity $\lambda_{p}$.
The clusters are of the form $N^{x_{i}}=N_{i}+x_{i}$ for each $x_{i}\in\phi_{p}$.
The $N_{i}$ are a family of identical and independently distributed
finite point sets with distribution independent of the parent process.
The complete process $\phi$ is given by \begin{eqnarray}
\phi & = & \bigcup_{x\in\phi_{p}}N^{x}.\end{eqnarray}
Note that the parent points themselves are not included. The daughter
points of the representative cluster $N_{0}$ are scattered independently
and with identical distribution $F(A)=\int_{A}f(x)\d x,\ A\subset\R^{2}$, 
around the origin. We also assume that the scattering density of the daughter process $f(x)$ is isotropic. This makes the process $\phi $ isotropic. The intensity of the cluster process is $\lambda=\lambda_{p}\bar{c}$,
where $\bar{c}$ is the average number of points in representative
cluster.

We further focus on more specific models for the representative cluster,
namely Matern cluster processes and Thomas cluster processes. In these
processes the number of points in the representative cluster is Poisson
distributed with mean $\bar{c}$. For the Matern cluster process each
point is uniformly distributed in a ball of radius $a$ around the
origin. So the density function $f(x)$ is given by \begin{eqnarray}
f(x) & = & \begin{cases}
\frac{1}{\pi a^{2}}, & \Vert x\Vert\leq a\\
0 & \text{otherwise}.\end{cases}\end{eqnarray}
In the Thomas cluster process each point is scattered using a symmetric
normal distribution with variance $\sigma^{2}$ around the origin.
So the density function $f(x)$ is given by\[
f(x)=\frac{1}{2\pi\sigma^{2}}\exp\left(-\frac{\Vert x\Vert^{2}}{2\sigma^{2}}\right).\]
\begin{figure}
\includegraphics[scale=0.55]{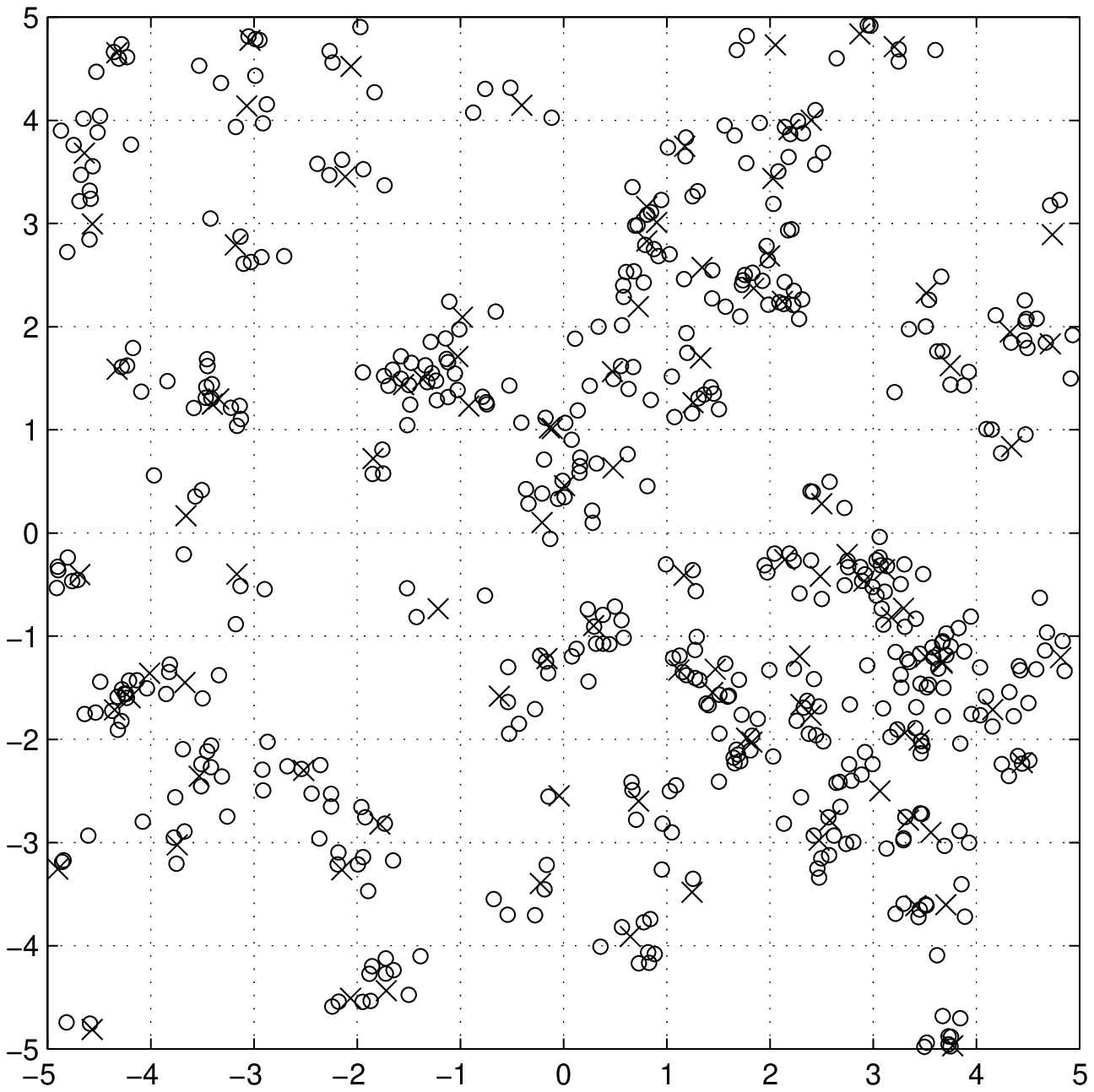}\includegraphics[scale=0.55]{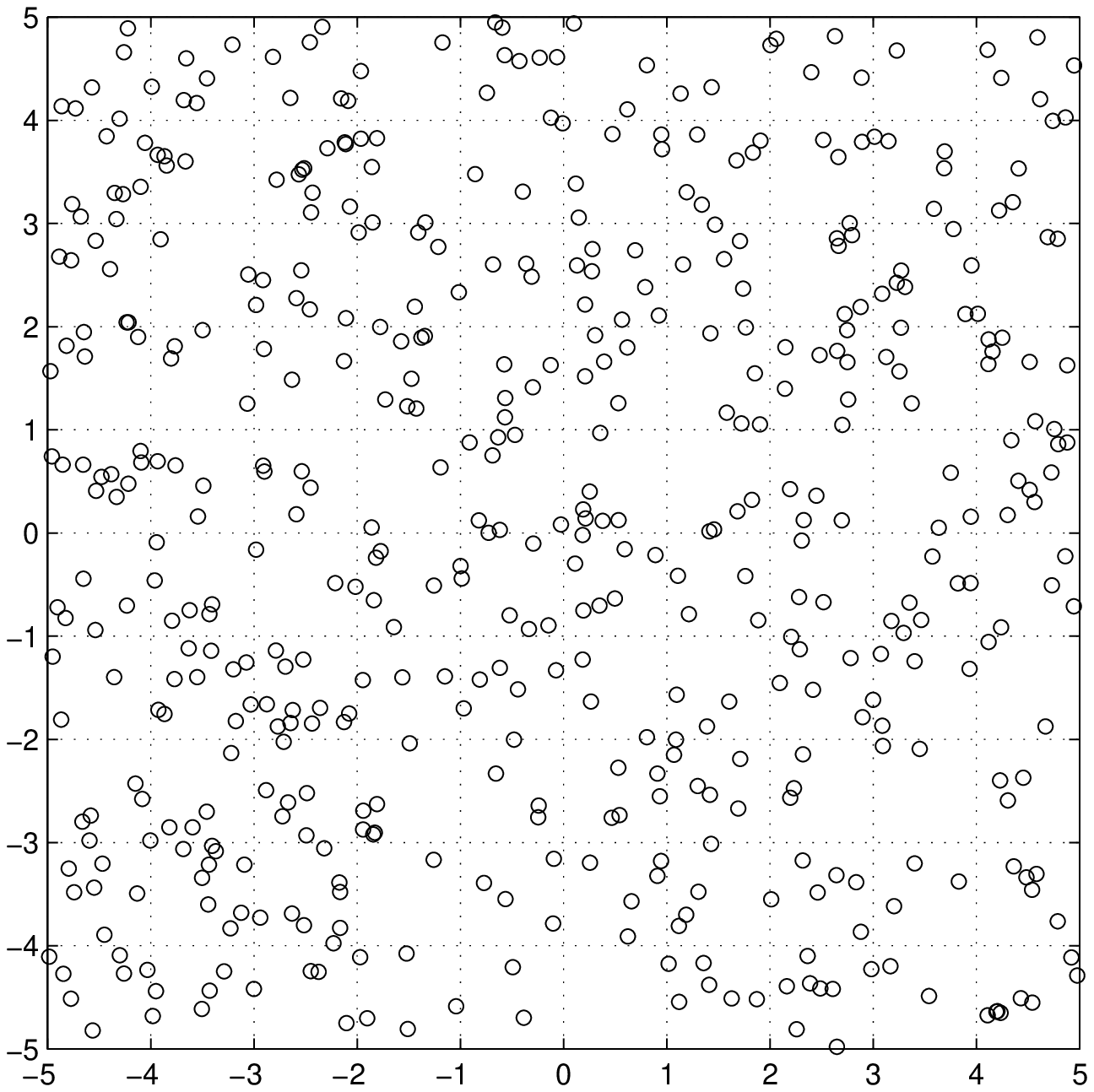}
\caption{(Left) Thomas cluster process with parameters $\lambda_{p}=1,\bar{c}=5$
and $\sigma=0.2$. The crosses indicate the parent points. (Right)
PPP with the same intensity $\lambda=5$ for comparison.}
\label{fig:illus1}
\end{figure}
A Thomas cluster process is illustrated in Fig.\ref{fig:fig1}.  Newman-Scott cluster processes
  are also a  Cox processes~\cite{stoyan} when the number of points in the daughter cluster are  Poisson distributed.  The density of the driving random measure   in this case is
\begin{equation*}
 \pi(y)=\bar{c}\sum_{x\in \phi_p}f(y-x)
\end{equation*}
Let
$E_{0}^{!}(.)$ denote the expectation with respect to the reduced
Palm measure \cite{stoyan,kallenberg}. It is basically the conditional
expectation for point processes, given the there is a point of the
process at the origin but without including the point. Let $v(x):\R^{2}\rightarrow[0,1]$
and $\int_{\R^2}\vert1-v(x)\vert \d x <\infty $. When $\phi$ is Poisson of intensity
$\lambda$, the conditional  generating functional is \begin{eqnarray}
E_{0}^{!}\left(\prod_{x\in\phi}v(x)\right) & = & E\left(\prod_{x\in\phi}v(x)\right)\nonumber\\
 & = & \exp\left(-\lambda\int_{\mathbb{R}^{2}}[1-v(x)]\d x\right)\end{eqnarray}
The generating functional $\tilde{G}(v)=E\left(\prod_{x\in\phi}v(x)\right)$
of  the Neyman-Scott cluster process is given by~\cite{stoyan,verejones}\[
\tilde{G}(v)=\exp\left( -\lambda_{p}\int_{\R^{2}}\left[1-M\left(\int_{\R^{2}}v(x+y)f(y)\d y\right)\right] \d x\right) \]
where $M(z)=\sum_{i=0}^{\infty}p_{n}z^{n}$ is the moment generating
function of the number of points in the representative cluster. When
the number of points in the representative cluster is Poisson with
mean $\bar{c}$, as in the case of Matern and Thomas cluster processes,
\[
M(z)=\exp(-\bar{c}(1-z)).\]
 The generating functional for the representative cluster $G_{c}(v)$
is given by \cite{verejones,cox} \[
G_{c}(v)=M\left(\int_{\R^{2}}v(x)f(x)\d x\right)\]
The reduced Palm distribution $P_{0}^{!}$ of a Neyman-Scott cluster
process $\phi$ is given by~\cite{stoyan,kallenberg,verejones,lothar:1986}
\begin{equation}
P_{0}^{!}=P*\tilde{\Omega}_{0}^{!}\label{palm1}\end{equation}
 where $P$ is the distribution of $\phi$, and $\tilde{\Omega}_{0}^{!}$
is the reduced Palm distribution of the finite representative cluster
process $N_{0}$. \char`\"{}$*$\char`\"{} denotes the convolution
of distributions, which corresponds to the superposition of $\phi$
and $N_{0}$. The reduced Palm distribution $\tilde{\Omega}_{0}^{!}$
is given by \begin{equation}
\tilde{\Omega}_{0}^{!}(Y)=\frac{1}{\bar{c}}E\Big(\sum_{x\in N_{0}}1_{Y}(\phi_{-x}\setminus\{0\})\Big)\label{eq:palm-one}\end{equation}
 where $\phi_{x}=\phi+x$,  is a  translated point process. We require
the following lemma to evaluate the conditional Laplace transform
of the interference. Let $\mathcal{G}(v)$ denote the conditional
generating functional of the Neyman-Scott cluster process, \ie,\begin{equation}
\mathcal{G}(v)=E_{0}^{!}\left(\prod_{x\in\phi}v(x)\right)\label{eq:mgf_cluster}\end{equation}
 We will use a dot to indicate the variable   which the functional
is acting on. For example $\mathcal{G}(v(\cdot -y))=E_{0}^{!}[\prod_{x\in\phi}v(x-y)]$.
\begin{lemma}
\label{lem:moment-gf} Let $0\leq v(x)\leq1$. The conditional generating
functional of Thomas and Matern clustered processes is \[
\mathcal{G}(v)=\tilde{G}(v)\int_{\R^{2}}G_{c}(v(\cdot -y))f(y)\d y.\]
\end{lemma}
\begin{proof}
Let $Y_{x}=Y+x$. From (\ref{eq:palm-one}), we have \begin{eqnarray}
\tilde{\Omega}_{0}^{!}(Y) & = & \frac{1}{\bar{c}}E\Big(\sum_{x\in N_{0}}1_{Y_{x}}(\phi\setminus\{x\})\Big)\end{eqnarray}
 Let $\Omega()$ denote the probability distribution of the representative
cluster. Using the Campbell-Mecke theorem~\cite{stoyan}, we get
\begin{eqnarray}
\tilde{\Omega}_{0}^{!}(Y) & = & \frac{1}{\bar{c}}\int_{\R^{2}}\int_{\mathcal{N}}1_{Y_{x}}(\phi)\Omega_{x}^{!}(\d \phi)\bar{c}F(\d x)\nonumber \\
 & = & \int_{\R^{2}}\int_{\mathcal{N}}1_{Y_{x}}(\phi)\Omega_{x}^{!}(\d \phi)f(x)\d x\label{eq:fixed1}\end{eqnarray}
 Here $\mathcal{N}$ denotes the space of locally finite and simple
point sequences~\cite{stoyan} on $\R^{2}$. Since the representative
cluster has a Poisson distribution of points, by Slivnyak's theorem
\cite{stoyan} we have $\Omega_{x}^{!}(.)=\Omega(.)$. Hence \begin{eqnarray}
\tilde{\Omega}_{0}^{!}(Y) & = & \int_{\R^{2}}\int_{N}1_{Y_{x}}(\phi)\Omega(\d \phi)f(x)\d x\nonumber \\
 & = & \int_{\R^{2}}\Omega(Y_{x})f(x)\d x\label{eq:silv}\end{eqnarray}
For notational convenience let $\psi$ denote $N_{0}$. Let $\psi_{y}=\psi+y$.
Using (\ref{palm1}), we have \begin{eqnarray}
\mathcal{G}(v) & = & \int_{\mathcal{N}}\int_{\mathcal{N}}\prod_{x\in\phi\cup\psi}v(x)P(\d \phi)\tilde{\Omega}_{0}^{!}(\d \psi)\nonumber \\
 & = & \int_{\mathcal{N}}\prod_{x\in\phi}v(x)P(\d \phi)\int_{\mathcal{N}}\prod_{x\in\psi}v(x)\tilde{\Omega}_{0}^{!}(\d \psi)\nonumber \\
 & = & \tilde{G}(v)\int_{\mathcal{N}}\prod_{x\in\psi}v(x)\tilde{\Omega}_{0}^{!}(\d \psi)\label{eq:fixed2}\\
 & \stackrel{(a)}{=} & \tilde{G}(v)\int_{\mathcal{N}}\prod_{x\in\psi}v(x)\int_{\R^{2}}\Omega(\d \psi_{y})f(y)\d y\nonumber \\
 & = & \tilde{G}(v)\int_{\R^{2}}\int_{\mathcal{N}}\prod_{x\in\psi}v(x)\Omega(\d \psi_{y})f(y)\d y\nonumber \\
 & = & \tilde{G}(v)\int_{\R^{2}}\int_{\mathcal{N}}\prod_{x\in\psi}v(x-y)\Omega(\d \psi)f(y)\d y\nonumber \\
 & \stackrel{(b)}{=} & \tilde{G}(v)\int_{\R^{2}}G_{c}(v(\cdot -y))f(y)\d y\nonumber \end{eqnarray}
 $(a)$ follows from (\ref{eq:silv}), and $(b)$ follows from the
definition of $\mathcal{G}(.)$. 
\end{proof}
So from the above lemma, we have \begin{eqnarray}
\mathcal{G}(v) & = & \exp\left(-\lambda_{p}\int_{\R^{2}}\Big[1-M\Big(\int_{\R^{2}}v(x+y)f(y)\d y\Big)\Big] \d x\right)\nonumber\\
 &  & \times\int_{\R^{2}}M\Big(\int_{\R^{2}}v(x-y)f(x)\d x\Big)f(y)\d y\end{eqnarray}
The above equation holds when all the integrals are finite.  Since
$f(x)=f(-x)$, then $\int_{\R^{2}}v(x+y)f(y)\d y=\int_{\R^{2}}v(x-y)f(y)\d y=v*f$,
so \begin{eqnarray}
\mathcal{G}(v) & = & \exp\left(-\lambda_{p}\int_{\R^{2}}\Big[1-M((v*f)(x))\Big]\d x\right)\int_{\R^{2}}M((v*f)(y))f(y)\d y\label{eq:laplace-full}\end{eqnarray}
Likelihood and nearest neighbor functions of the Poisson cluster process,
which involve similar calculations with Palm distributions are provided
in \cite{baudin1981lan}. One can obtain the nearest-neighbor distribution
function of Thomas or Matern cluster process as $D(r)=\mathcal{G}(1_{B(o,r)^{c}}(.))$. In some cases the number of points per cluster may be fixed rather than Poisson. 
The conditional generating functional, for this case  is given in Appendix \ref{sub:fixed}.

\section{Interference and Outage Probability of Poisson Cluster Processes\label{sec:Outage-Probability}}

In this section, we first derive the characteristics of interference
in a Poisson clustered process conditioned on the existence of a transmitting
node at the origin. We then evaluate the outage probability for a
transmit-receive pair when the transmitters are distributed as a Neyman-Scott
cluster process, with the number of points in each cluster is  Poisson
with mean $\bar{c}$ and density function $f(x)$. 
\subsection{Properties of the Interference $I_{\phi}(z)$}
Let $\mathcal{L}_{h }(s)$ denote the Laplace transform of  the  fading random variable $h  $. 
\begin{lemma}
\label{lem:laplace-interference} The conditional Laplace transform
of the interference is given by
\begin{equation}
\mathcal{L}_{I_{\phi}(z)}(s)=\mathcal{G}\left(\mathcal{L}_h (sg(\cdot -z))\right)
\label{lap1} 
\end{equation}
\end{lemma}
 
\begin{proof}
From (\ref{eq:interfernce-main}) we have \begin{eqnarray}
\mathcal{L}_{I_{\phi}(z)}(s) & = & E_{0}^{!}\exp(-s\sum_{x \in\phi}h_xg(x -z))\nonumber\\
 & = & E_{0}^{!}\left[\prod_{x \in\phi}\exp(-sh_x g(x -z))\right]\nonumber\\
 & \stackrel{(a)}{=} & E_{0}^{!}\left[\prod_{x \in\phi}\mathcal{L}_{h }(sg(x -z))\right]\end{eqnarray}
where $(a)$ follows from the independence of $h_{x}$ and (\ref{lap1})
follows from (\ref{eq:mgf_cluster}). 
\end{proof}
We observe from Lemma \ref{lem:laplace-interference} and \eqref{eq:laplace-full},
that the conditional Laplace transform of the interference $\mathcal{L}_{I_{\phi}(z)}(s)$
depends on the position $z$. This implies that the distribution of
the interference depends on the location $z$  at which
we observe the interference. This is in contrast to the fact that
the interference distribution is independent of the location $z$
when the transmitters are Poisson distributed on the plane \cite{bacelli-aloha,WebAnd2006a}.
This is due to the non-stationarity of the reduced Palm measure of
the Neyman-Scott cluster processes.   If one interprets $I_\phi(z)$ as a stochastic process, it is then  a non stationary process  due to the above reason.

Let $\mathcal{K}_{n}(B)$ denote the reduced $n$-th factorial moment
measure \cite{stoyan,verejones} of a point process $\psi$, and let
$B=B_{1}\times\ldots\times B_{n-1},\ B_{i}\in\mathbb{R}^{2}$.\begin{eqnarray}
\mathcal{K}_{n}(B) & = & E_{0}^{!}\left[\sum_{x_{1},\ldots,x_{n-1}\in\psi}^{x_{i}\neq x_{j}}1_{B}(x_{1},\ldots,x_{n-1})\right]\end{eqnarray}
$\mathcal{K}_{2}(B(0,R))$, for example, denotes the average number
of points inside a ball of radius $R$ centered around the origin,
given that a point exists at the origin.   First and second moments
of the interference can be determined using the second and third order
reduced factorial moments. The average interference (conditioned on
the event that there is a point of the process at the origin) is given
by\begin{eqnarray}
E_{0}^{!}[I_{\phi}(z)] & = & E_{0}^{!}\left[\sum_{x\in\phi}h_xg(x-z)\right]\nonumber\\
 & = & E[h]\lambda\int_{\R^{2}}g(x-z)\mathcal{K}_{2}(\d x)\end{eqnarray}
Since the process $\phi$  is stationary, $\mathcal{K}_{2}(B)$ can be expressed
as \cite{stoyan,hanisch-1983}\[
\mathcal{K}_{2}(B)=\frac{1}{\lambda^{2}}\int_{B}\rho^{(2)}(x)\d x,\]
where $\rho^{(2)}(x)$ is the second order product density%
\footnote{Intuitively, this indicates the probability that there are two points
separated by $\Vert x\Vert$. For PPP, it is $\rho^{(2)}(x)=\lambda^{2}$
independent of $x$. Also the second order product density is a function
of two arguments \ie, $\rho^{(2)}(x_{1,},x_{2})$. But when the process
$\phi$ is stationary, $\rho^{(2)}$ depends only on the difference
of its arguments \ie, $\rho^{(2)}(x_{1},x_{2})=\nu(x_{1}-x_{2})$
for all $x_{1},x_{2}\in\mathbb{R}^{2}$. Furthermore if $\phi$ is
motion-invariant, i.e., stationary and isotropic, then $\nu$ depends
only on $\Vert x_{1}-x_{2}\Vert$ \cite[pg 112]{stoyan}.%
}.\emph{ }So we have\emph{ \begin{equation}
E_{0}^{!}[I_{\phi}(z)]=\frac{E[h]}{\lambda}\int_{\R^{2}}g(x-z)\rho^{(2)}(x)\d x \label{eq:mean-interference}\end{equation}
Example: Thomas Cluster Process}. In this case, from \cite{stoyan}
\[
\frac{\rho^{(2)}(x)}{\lambda^{2}}=1+\frac{1}{4\pi\lambda_{p}\sigma^{2}}\exp\Big(\frac{-\Vert x\Vert^{2}}{4\sigma^{2}}\Big)\]
 where $\lambda=\lambda_{p}\bar{c}$. We obtain \begin{equation}
E_{0}^{!}[I_{\phi}(z)]=EI_{\text{Poi }(\lambda)}+\frac{\bar{c}E[h ]}{4\pi\sigma^{2}}\int_{\R^{2}}g(x-z)\exp\left(\frac{-\Vert x\Vert^{2}}{4\sigma^{2}}\right)\d x\label{eq:average}\end{equation}

Where $EI_{\text{Poi }(\lambda)}$ is the average interference seen
by a receiver located at $z$, when the nodes are distributed as a
PPP with intensity $\lambda$. The above expression also  shows that
the mean interference%
\footnote{Note that for $g(x)=\Vert x\Vert^{-\alpha},\ E_{0}^{!}[I_{\phi}(z)]$
is diverging.%
} is indeed larger than for the PPP. One can also get the above from
the conditional Laplace transform in Lemma \ref{lem:laplace-interference}
and using $E_{0}^{!}[I_{\phi}(z)]=-\frac{d}{ds}\mathcal{L}_{I_{\phi}(z)}(s)|_{s=0}$.
In the following theorem we provide bounds to the tail probability of
the interference $I_{\phi}(z)$ for {\em any  stationary distribution $\phi$} of transmitters. We adapt the technique presented
in \cite{WebAnd2006d} to derive the tail bounds of the interference.
We denote the tail probability (CCDF) of $I_{\phi}(z)$ by $\bar{F}_{I}(y)=\P(I_{\phi}(z)\geq y)$. 
 
\begin{thm}
\label{thm:CCDF}When the transmitters are distributed as a  stationary and  isotropic  point process $\phi$  of intensity $\lambda $ with conditional generating functional $\mathcal{G}$ and second order product density $\rho^{(2)} 	$, the tail probability $\bar{F}_{I}(y)$ of the interference
at location $z$, conditioned on a transmitter present at the origin%
\footnote{We do not include the contribution of the transmitter at the origin in
the interference. This is because the transmitter at the origin is
the intended transmitter which we  focus  on.%
} is lower bounded by  $\bar{F}_{I}^{l}(y)$
and upper bounded by $\bar{F}_{I}^{u}(y)$, where  
\begin{equation}
 \bar{F}_{I}^{l}(y)=1-\mathcal{G}\left(F_{h }\left(\frac{y}{g(.-z)}\right)\right)
\end{equation}
 \begin{equation}
 \bar{F}_{I}^{u}(y)=1-(1-\varphi(y))\mathcal{G}\left(F_{h}\left(\frac{y}{g(.-z)}\right)\right)
\end{equation}
where $F_{h }(x)$ denotes the CDF of the power  fading
coefficient $h $ and \[
\varphi(y)=\frac{1}{y\lambda}\int_{\R^{2}}g(x-z)\rho^{(2)}(x)\int_{0}^{y/g(x-z)}\nu \d F_{h }(\nu)\d x.\]

\end{thm}
\begin{proof}
The basic idea is to divide the transmitter set into two subsets $\phi_{y}$ and
$\phi_{y}^{c}$ where, \begin{eqnarray}
\phi_{y} & = & \{x\in\phi,\ h_{x} g(x-z)>y\}\\
\phi_{y}^{c} & = & \{x\in\phi,\ h_{x} g(x-z)\leq y\}\end{eqnarray}
 $\phi_{y}$ consists of those transmitters,  whose contribution     to the interference exceeds $y$. We have $I_{\phi}(z)=I_{\phi_{y}}(z)+I_{\phi_{y}^{c}}(z)$,
where $I_{\phi_{y}}(z)$ corresponds to the interference due to the
transmitter set $\phi_{y}$ and $I_{\phi_{y}^{c}}(z)$ corresponds
to the interference due to the transmitter set $\phi_{y}^{c}$. Hence
we have \begin{eqnarray}
\bar{F}_{I}(y) & = & \P(I_{\phi_{y}}(z)+I_{\phi_{y}^{c}}(z)\geq y)\nonumber\\
 & \geq & \P(I_{\phi_{y}}(z)\geq y)\nonumber\\
 & = & 1-\P(I_{\phi_{y}}(z)<y)\nonumber\\
 & = & 1-\P(\phi_{y}=\emptyset).\end{eqnarray}
We can evaluate the probability $\P(\phi_{y}=\emptyset)$ that $\phi_{y}$
is empty  using the conditional Laplace functional as follows:
\begin{eqnarray}
 \P(\phi_{y}=\emptyset) & = & E_{0}^{!}\prod_{x\in\phi}  1_{h_{x} g(x-z)\leq y} \nonumber\\ 
 & \stackrel{(a)}{=} & E_{0}^{!}\prod_{x\in\phi}  E_{h_{x} }\left(1_{h_{x} g(x-z)\leq y}\right)\nonumber \\
 & = & E_{0}^{!}\prod_{x\in\phi}F_{h }\left(\frac{y}{g(x-z)}\right)\nonumber\\
 & = & \mathcal{G}\left(F_{h }\left(\frac{y}{g(\cdot-z)}\right)\right),\end{eqnarray}
where $(a)$ follows from the independence of $h_x$. 
To obtain the upper bound \begin{eqnarray}
\bar{F}_{I}(y) & = & \P(I_{\phi}>y \:| \:I_{\phi_{y}}>y)\bar{F}_{I}^{l}(y)+\P(I_{\phi}>y\:| \:I_{\phi_{y}}\leq y)(1-\bar{F}_{I}^{l}(y))\nonumber\\
 & \stackrel{(a)}{=} & 1-\mathcal{G}\left(F_{h }\left(\frac{y}{g(\cdot-z)}\right)\right)+\P(I_{\phi}>y\:| \:I_{\phi_{y}}\leq y)\mathcal{G}\left(F_{h }\left(\frac{y}{g(\cdot -z)}\right)\right)\nonumber\\
 & = & 1-(1-\P(I_{\phi}>y\:| \:I_{\phi_{y}}\leq y))\mathcal{G}\left(F_{h }\left(\frac{y}{g(\cdot -z)}\right)\right)\end{eqnarray}
 where $(a)$ follows from the lower bound we have established. To
evaluate $\P(I_{\phi}>y\:| \:I_{\phi_{y}}\leq y)$ we use the Markov inequality
(the 	Chebeshev inequality can also be used but is more difficult to be
evaluated in this particular setting). We have \begin{eqnarray}
\P(I_{\phi}>y\:|\:I_{\phi_{y}}\leq y) & = & \P(I_{\phi}>y\:|\:\phi_{y}=\emptyset)\nonumber\\
 & \stackrel{(a)}{\leq} & \frac{E_{0}^{!}\left(I_{\phi}\:|\:\phi_{y}=\emptyset\right)}{y}\nonumber\\
 & = & \frac{1}{y}E_{0}^{!}\sum_{x\in\phi}h_{x} g(x-z)1_{h_{x} g(x-z)\leq y}\nonumber\\
 & = & \frac{1}{y}E_{0}^{!}\sum_{x\in\phi}g(x-z)\int_{0}^{y/g(x-z)}\nu \d F_{h }(\nu)\nonumber\\
 & \stackrel{(b)}{=} & \frac{1}{y\lambda}\int_{\R^{2}}g(x-z)\int_{0}^{y/g(x-z)}\nu \d F_{h }(\nu)\rho^{(2)}(x)\d x\end{eqnarray}
 $(a)$ follows from the Markov inequality,  and $(b)$ follows from a procedure
similar to the calculation of the mean interference in \eqref{eq:mean-interference}. 
\end{proof}
In the proof of Lemma \ref{lem:scaling}, we show $\varphi(y)\sim\theta_{2}y^{-2/\alpha}$
when $g(x)=\Vert x\Vert^{-\alpha}$. This indicates the tightness
of the bounds for large $y$. Lemma \ref{lem:scaling} shows that the interference
is a heavy-tailed distribution with parameter $2/\alpha$ when the nodes are distributed as a Neyman-Scott cluster process. 

\begin{lemma}
\label{lem:scaling}For $g(x)=\Vert x\Vert^{-\alpha}$, the lower
and upper bounds to CCDF $\bar{F}_{I}(y)$ of the interference at location $z$, when the nodes are distributed as a Neyman-Scott cluster process scale as follows for $y\rightarrow\infty$.\begin{eqnarray}
\bar{F}_{I}^{l}(y) & \sim & \theta_{1}y^{-2/\alpha}\\
\bar{F}_{I}^{u}(y) & \sim & (\theta_{1}+\theta_{2})y^{-2/\alpha}\end{eqnarray}
where $\theta_{1}=\pi\bar{c}[(f*f)(z)+\lambda_{p}]\int_{0}^{\infty}\nu^{2/\alpha}\d F_{h }(\nu)$
and $\theta_{2}=2\theta_{1}/(\alpha-2)$. 
\end{lemma}
\begin{proof}
See Appendix \ref{sub:Proof-of-Lemma-scaling}. 
\end{proof}
Remarks:
\begin{enumerate}
 \item Observe that $\theta_1=\pi\lambda^{-1} \rho^{(2)}(z)\int_{0}^{\infty}\nu^{2/\alpha}\d F_{h }(\nu)$. A similar kind of scaling law  with  $\theta_1=\pi\lambda^{-1} \rho^{(2)}(z)E_h[\nu^{2/\alpha}]$ and $\theta_2=2\theta_1/(\alpha-2)$ can be obtained  when the transmitters are scattered as any ``nice''\footnote{ We require the conditional  generating functional to have a series expansion with respect to  reduced  $n$-th factorial moment measures of the reduced Palm distribution~\cite{hanisch-1983} similar to that of the expansion of 
 generating functional ~\cite[p.116]{stoyan} and \cite{westcott-1972}.   The proof  of the existence and the series expansion of the conditional  generating functional  with respect to  reduced  $n$-th factorial moment measures, would be of more technical nature     following   a technique used in ~\cite{westcott-1972}. If such an expansion exists it is straightforward to prove    the scaling laws for the CCDF of  interference similar to Lemma~\ref{lem:scaling},    with $\theta_1=\pi\lambda^{-1} \rho^{(2)}(z)E_h[\nu^{2/\alpha}]$ and $\theta_2=2\theta_1/(\alpha-2)$.  }  stationary, isotropic  point process with intensity $\lambda$ and second order product density $\rho^{(2)}(x)\neq 0$ at $x=z$. 
\item A similar heavy-tailed distribution with parameter $2/\alpha$ was obtained
for Poisson interference in \cite{Sousa90,WebAnd2006d}. Since $2/\alpha<1$,
the mean and hence the variance diverge. This can also be inferred
from \eqref{eq:average} and is  due to the singularity of the channel
function $g(x)=\Vert x\Vert^{-\alpha}$ at the origin.  For Matern cluster processes $(f*f)(z)=0$, for $\Vert z \Vert >2a$ and for Thomas cluster processes  $(f*f)(z)$ is a  Gaussian  with variance $2\sigma^2$.  Hence for large $z$, we observe that the constants $\theta_1$ become similar to that
of the unconditional interference. This is because, the contribution of the cluster at origin becomes small as we move far from the origin.
\item When the path
loss function is $g(x)=(1+\Vert x\Vert^{\alpha})^{-1}$, the distribution
of the interference   more  strongly depends  on the fading model. Using a similar
proof as in  Lemma \ref{lem:scaling}, one can deduce an exponential
tail decay when $g(x)=(1+\Vert x\Vert^{\alpha})^{-1}$ and Rayleigh fading. 
\begin{figure}[h]
\begin{centering}
\includegraphics[scale=0.6, ]{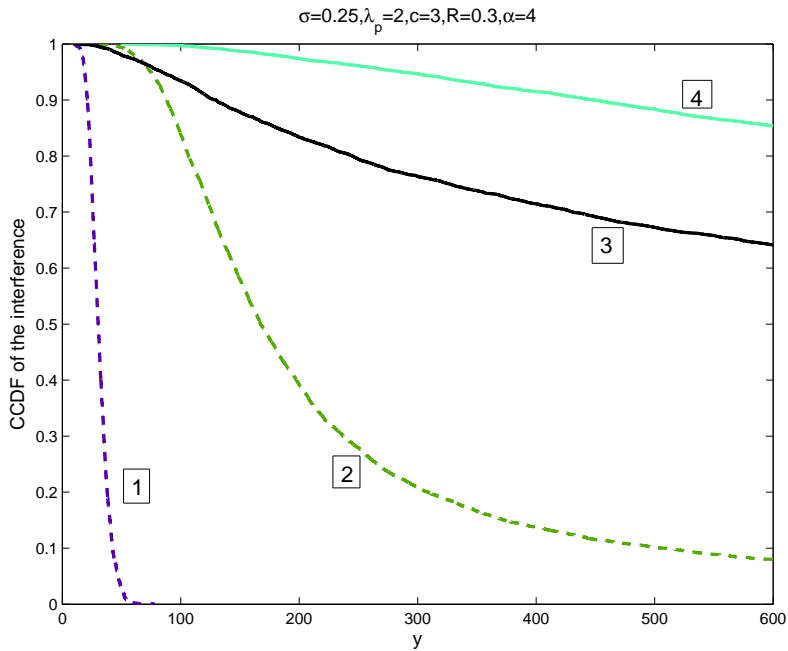}
\par\end{centering}
 \caption{ $\lambda_p=2,\bar{c}=3,\sigma=0.25,\alpha=4,R=0.3$: Comparison of   the  interference CCDF for different path-loss models and  different  fading. They were generated using Monte-Carlo simulation.  Curves $\#1$  and $\#2$ correspond to  $g(x)=(1+\Vert x\Vert^{\alpha})^{-1}$. Curve $\#1$ corresponds to Rayleigh fading and exhibits an exponential decay. Curve $\#2$  for which  $h$ is distributed as generalized Pareto with parameters $k=1,\theta=0,\sigma_p=1$ (a hypothetical power  fading distribution which exhibits power law decay) exhibits a power law decay.   Curves $\#3$ (generalized Pareto )  and $\#4$ (Rayleigh) correspond to   $g(x)=\Vert x\Vert^{-\alpha}$ and exhibit a heavy tail  for both fading distributions.}
\label{fig:comparison}
\end{figure}
Similarly  if the  power fading coefficient follows a  power-law distribution with exponent $k$, the
tail of the interference shows a power-law decay. This is because
of the presence of the term $y^{-2/\alpha}\int_{y}^{\infty}[1-F_{h }(u)](u-y)^{2/\alpha-1}\d u$
in the proof. So when using non-singular channel models, the interference
has a more intricate dependence on the fading characteristics rather
than a simple dependence on $E_h[\nu^{2/\alpha}]$
 as in the singular case.   This   behavior is  well  understood for  Poisson  and  unconditional Poisson cluster shot noise process~\cite{lowen1990pls,ty-tail}.  The   properties of  interference for different path loss models  with  no fading, when  the nodes are uniformly distributed  are discussed in~\cite{Inaltekin}.
\end{enumerate}

\subsection{Success probability: $\mathbb{P}(\text{success})$}
Let the desired transmitter be located at the origin and the receiver
at location $z$ at distance $R=\|z\|$ from the transmitter. With
a slight abuse of notation we shall be using $R$ to denote the point
$(R,0)$. The probability of success for this pair is given by \begin{equation}
\P(\text{success})=\P^{!0}\Big(\frac{h  g(z)}{W+I_{\phi}(z)}\geq T\Big)\label{eq:main}\end{equation}
 We now assume Rayleigh fading, \ie, the received power is exponentially
distributed with  mean $\mu$. So we have \begin{align}
\P(\text{success}) & =\int_{0}^{\infty}e^{-\mu sT/g(z)}\d \P(W+I_{\phi\setminus\{0\}}(z)\leq s)\nonumber \\
 & =\mathcal{L}_{I_{\phi}(z)}(\mu T/g(z))\mathcal{L}_{W}(\mu T/g(z))\,,\label{main-outage}\end{align}
 When $h_x $ is Rayleigh we have \begin{eqnarray}
\mathcal{L}_{h }(sg(x -z)) & = & \frac{\mu}{\mu+sg(x -z)}\label{eq:rayleigh}\end{eqnarray}
 At $s=\mu T/g(R)$ we observe that the above expression will be independent
of the mean of the exponential distribution $\mu$.

\begin{lemma}
\label{lem:outage0}{[}Success probability] The probability of successful
transmission between the transmitter at the origin and the receiver
located at $z\in\mathbb{R}^{2}$, when $W\equiv0$ (no noise), is
given by \begin{align}
\P(\text{success})= & \underbrace{\exp\Big\{-\lambda_{p}\int_{\R^{2}}\Big[1-\exp(-\bar{c}\beta(z,y))\Big]\d y\Big\}}_{T_{1}}\nonumber \\
 & \times\underbrace{\int_{\R^{2}}\exp(-\bar{c}\beta(z,y))f(y)\d y}_{T_{2}}\label{eq:main}\end{align}
 where \begin{equation}
\beta(z,y)=\int_{\R^{2}}\frac{g(x-y-z)}{\frac{g(z)}{T}+g(x-y-z)}f(x)\d x\end{equation}

\end{lemma}
\begin{proof}
Follows from \eqref{eq:rayleigh} and Lemma\foreignlanguage{american}{\emph{
}}\ref{lem:laplace-interference}.
\end{proof}
The success probability, when the number of nodes in each cluster
is fixed is given in the Appendix \ref{sub:fixed}. See Figure \ref{fig:comparison}
for comparison.
\begin{figure}[h]
\begin{centering}
\includegraphics[scale=0.5, ]{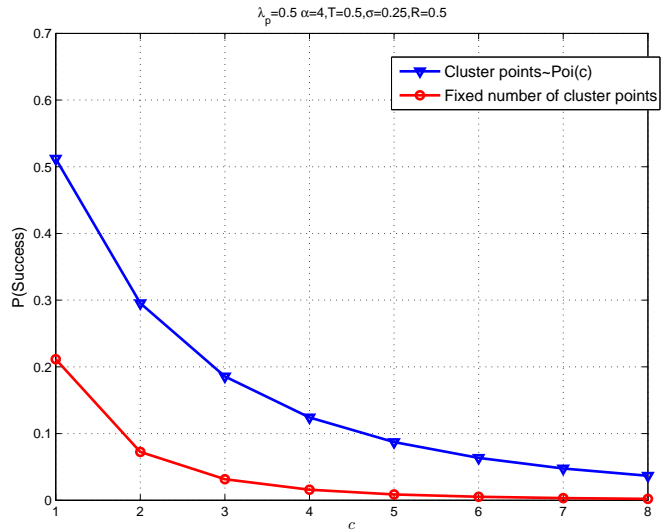}
\par\end{centering}
\caption{Comparison of $\mathbb{P}(\text{success)}$ when the number of points
in a cluster are fixed and Poisson distributed with parameter $\bar{c}$.}
\label{fig:comparison}
\end{figure}
When the fading is Nakagami-$m$, the probability of success is evaluated
in the Appendix \ref{sub:nakagami}  for integer  $m$.   

{\em Remarks:}
\begin{enumerate}
\item The term $T_{1}$ in \eqref{eq:main} captures the interference without
the cluster at the origin (\ie, without conditioning); it is independent%
\footnote{By this we mean the unconditional interference distribution which
leads to this term does not depend on the location $z$. The term
$T_{1}$ does depend on $g(z)$.%
} of the position $z$ since the original cluster process is stationary
(can be verified by change of variables $y_{1}=y+z$). The second
term $T_{2}$ is the contribution of the transmitter's cluster; it
is identical for all $z$ with $\|z\|=R$ since $f$ and $g$ are
isotropic. So the success probability itself is the same for all $z$
at distance $R$. This is because the Palm distribution is always
 isotropic  when the original distribution is
motion-invariant~\cite{stoyan}.  Hence  we shall use $\beta(R,y)$ to denote $\beta(z,y)$ where $z=Re^{i\theta}$. We shall also use $R$ and $(R,0)$ interchangeably and will be clear by the context.
 
\item From the above argument we observe that $\P(\text{success})$ depends
only on $\ \|z\|=R$ and not on the angle of $z$. So the success
probability should be interpreted as an average over the circle  $\|z\|=R$,
\ie, the receiver may be uniformly located anywhere on the circle
of radius $R$ around the origin. For large distances $R$, there is
a very high probability that the receiver is located in an empty space
and not in any cluster. Hence for large $R$ the  success probability 
is higher than that of a PPP of the same intensity. If the receiver
is also conditioned to be in a cluster, we have to multiply (at least
heuristically) by a term that is similar
to $T_{2}$ and this would significantly reduce the success probability.
\item From Lemma \ref{lem:outage0}, we have $\mathbb{P}(\text{success})=E_{0}^{!}[\exp(-sI_{\phi}(z))]$
evaluated at $s=\mu T/g(z)$. If $\mu T/g(z)$ is \emph{small, }and
\emph{$\int g(z)\d z<\infty$ (i.e, }finite average interference)\emph{
}then, $\mathbb{P}(\text{success})\leq P_{p}(\lambda)$. This follows
from \eqref{eq:average} and the fact that $E_{0}^{!}[I_{\phi}(z)]$
is the slope of the curve $E_{0}^{!}[\exp(-sI_{\phi}(z))]$ at $s=0$.
This implies that at small distances, spread spectrum (DS-CDMA) works better with a Poisson
distribution of nodes. (If the distance $R$
is large, then the spreading gain has to increase approximately like
$g(z)$  to keep $\mu T/g(z)$ small.)
\item Let $M$ be the DS-CDMA spreading factor. We have $\P(\text{outage})=\P(I_{\phi}(z)>\frac{M}{T}h_{x} g(z))$.
For $g(x)=\Vert x\Vert^{-\alpha}$, we have the following scaling
law for the outage probability with respect to the spreading gain.	\begin{eqnarray}
\theta_{1}R^{2}M^{-2/\alpha}T^{2/\alpha}E_{h }[\nu^{-2/\alpha}]\stackrel{(a)}{\lesssim} & \P(\text{outage}) & \stackrel{(b)}{\lesssim}\frac{\alpha}{\alpha-2}\theta_{1}R^{2}M^{-2/\alpha}T^{2/\alpha}E_{h }[\nu^{-2/\alpha}],\end{eqnarray}
where $E_{h }[\nu^{-2/\alpha}]=\int_{0}^{\infty}\nu^{-2/\alpha}\d F_{h^{2}}(\nu)$.
$(a)$ and $(b)$ follow from Lemma \ref{lem:scaling}. Also observe
that these scaling bounds are valid for any fading distribution for
which $E_{h }[\nu^{-2/\alpha}]<\infty$. Similar scaling laws with
the exponent of $M$ being $-2/\alpha$ can be obtained  when the
transmitters are Poisson distributed on the plane. When the  fading is Rayleigh \ie, $h\sim\exp(\mu)$, the lower bound is \[\pi\bar{c}[(f*f)(R)+\lambda_p]\Gamma\Big(1+\frac{2}{\alpha}\Big)\Gamma\Big(1-\frac{2}{\alpha}\Big) R^{2}M^{-2/\alpha}T^{2/\alpha}\] and the upper bound is $\alpha/(\alpha-2)$ times the lower bound. $\Gamma(z)$ represents the  standard Gamma function.
\end{enumerate}
 
We now derive closed form upper and lower bounds on $\mathbb{P}(\text{success})$.

\begin{lemma}
{[}Lower bound] \label{lem:lower} \begin{eqnarray}
\P(\text{success})\geq P_{p}(\lambda)P_{p}(\bar{c}\hat{f}^{*})\end{eqnarray}
where $P_{p}(\lambda)$ denotes the success probability when $\phi$
is a PPP, $\hat{f}^{*}=\sup_{y\in\R^{2}}(f*f)(y)$, and $\lambda=\lambda_{p}\bar{c}$. 
\end{lemma}
\begin{proof}
The first factor in (\ref{eq:main}), $T_1$ can be lower bounded by the success
probability in the standard PPP $P_{p}(\lambda)$, and the second
factor can be lower bounded by $P_{p}(\bar{c}\hat{f}^{*})$. From
\eqref{eq:main} and the fact that $\ 1-\exp(-\delta x)\leq\delta x,\delta\geq0$,
we have \begin{eqnarray}
\P(\text{success}) & \geq & \underbrace{\exp\Big(-\lambda_{p}\bar{c}\int_{\R^{2}}\beta(R,y)\d y\Big)}_{\text{Term1}}\\
 &  & \times\underbrace{\int_{\R^{2}}\exp(-\bar{c}\beta(R,y))f(y)\d y}_{\text{Term2}}\nonumber\end{eqnarray}
 \begin{eqnarray}
\text{Term1} & = & \exp\Big(-\lambda\int_{\R^{2}}\beta(R,y)\d y\Big)\nonumber\\
 & \stackrel{(a)}{=} & \exp\Big(-\lambda\int_{\R^{2}}\frac{g(y)}{\frac{g(R)}{T}+g(y)}\d y\Big)\nonumber\\
 & = & P_{p}(\lambda)\end{eqnarray}
 $(a)$ follows from change of variables, interchanging integrals
and using $\int f(x)=1$. \begin{eqnarray}
\text{Term2} & = & \int_{\R^{2}}\exp(-\bar{c}\beta(R,y))f(y)\d y\nonumber\end{eqnarray}
 Since $\exp(-x)$ is convex and $f(x)>0,\int f(x)=1$, Using Jensen's
inequality ($Ef(x)\geq f(E(x))$) we have, \begin{eqnarray}
\text{Term2} & \geq & \exp\Big(-\bar{c}\int_{\R^{2}}\beta(R,y)f(y)\d y\ \Big)\nonumber\end{eqnarray}
 Changing variables and using $f(x)=f(-x)$,we get, \begin{eqnarray}
\text{Term2} & \geq & \exp\Big(-\bar{c}\int_{\R^{2}}\frac{g(x)}{\frac{g(R)}{T}+g(x)}\int_{\R^{2}}f(x+z-y)f(y)\d y\d x\ \Big)\nonumber\\
 & \geq & \exp\Big(-\bar{c}\int_{\R^{2}}\frac{g(x)}{\frac{g(R)}{T}+g(x)}(f*f)(x+z)\d x\ \Big)\end{eqnarray}
 Hence \begin{eqnarray}
\text{Term2} & \geq & P_{p}(\bar{c}\hat{f}^{*})\end{eqnarray}
\end{proof}
Since $f\in L_{p}\ $, by Young's inequality \cite{folland} we have
$\hat{f}^{*}\leq\Vert f\Vert_{p}\Vert f\Vert_{q}$, where $1/p+1/q=1$
(conjugate exponents). For $a\geq1/\sqrt{\pi}$ (Matern) and $\sigma\geq1/\sqrt{2\pi}$
(Thomas), we get $\P(\text{success})\geq P_{p}(\lambda)P_{p}(\bar{c})$.
In general, $\hat{f}^{*}\leq\Vert f\Vert_{\infty}\Vert f\Vert_{1}$,
which is $1/\pi a^{2}$ for Matern and $1/2\pi\sigma^{2}$ for Thomas
processes. In the latter case, when $f$ is Gaussian, $f*f$ is also
Gaussian with variance $2\sigma^{2}$, hence $\hat{f}^{*}\leq1/4\pi\sigma^{2}$.
From \cite{bacelli-aloha}, we get  (by change of variables):\begin{equation}
P_{p}(\lambda)=\exp\left(-\lambda\int_{\mathbb{R}^{2}}\beta(R,y)\d y\right).\label{eq:poisson}\end{equation}
We have 
\begin{itemize}
\item for $g(x)=\Vert x\Vert^{-\alpha}$, $P_{p}(\lambda)=\exp(-\lambda R^{2}T^{2/\alpha}C(\alpha))$\cite{bacelli-aloha},
where $C(\alpha)=\big(2\pi\Gamma(2/\alpha)\Gamma(1-2/\alpha)\big)/\alpha=\frac{2\pi^{2}}{\alpha}\csc(2\pi/\alpha)$.
\item for $g(x)=(1+\Vert x\Vert^{\alpha})^{-1}$, $P_{p}(\lambda)=\exp(-\lambda TC(\alpha)(T+g(R))^{2/\alpha-1}g(R)^{-2/\alpha})$.
\end{itemize}
Let $\beta_{I}=\int_{\R^{2}}\beta(R,y)\d y$, $\hat{\beta}=\sup_{y\in\R^{2}}\beta(R,y)$
and $\hat{f}=\sup_{y\in\R^{2}}f(y)$. By H\"olders inequality we have
$\hat{\beta}\leq\min\{1,\hat{f}\beta_{I}(R)\}$. Also let $\kappa=\int_{\R^{2}}\beta(R,y)f(y)\d y$.

\begin{lemma}
{[}Upper bound]\label{lem:[Upper-bound] } \begin{eqnarray}
\P(\text{success}) & \leq & P_{p}\left(\frac{\lambda}{1+\bar{c}\hat{\beta}}\right)\end{eqnarray}
 
\end{lemma}
\begin{proof}
Neglecting the second term $T_{2}$ and using $\exp(-\delta x)\leq1/(1+\delta x)$,
we have\begin{eqnarray}
\mathbb{P}(\text{success}) & \leq & \exp\Big(-\lambda_{p}\int_{\R^{2}}\Big[1-\frac{1}{1+\bar{c}\beta(R,y)}\Big]\d y\Big)\nonumber\\
 & = & \exp\Big(-\lambda_{p}\int_{\R^{2}}\frac{\bar{c}\beta(R,y)}{1+\bar{c}\beta(R,y)}\d y\Big)\nonumber\\
 & \leq & \exp\Big(-\frac{\lambda_{p}\bar{c}}{1+\bar{c}\hat{\beta}}\int_{\R^{2}}\beta(R,y)\d y\Big)\end{eqnarray}

\end{proof}
From the above two lemmata, we get \begin{eqnarray}
P_{p}(\lambda)P_{p}(\bar{c}\hat{f}^{*})\leq & \P(\text{success}) & \leq P_{p}\left(\frac{\lambda}{1+\bar{c}\hat{\beta}}\right)\end{eqnarray}
from which follows $\P(\text{success})\rightarrow P_{p}(\lambda)$
as $\frac{\bar{c}}{\sigma},\frac{\bar{c}}{a}\rightarrow0$ as expected.
In Lemma \ref{lem:[Upper-bound] }, we have neglected the contribution
of the transmitter's cluster. We derive the following upper bound
in the proof of Lemma \ref{lemma4}, \begin{eqnarray}
\P(\text{success}) & \leq & P_{p}(\lambda)\exp\left(\lambda\beta_{I}\nu(\bar{c}\hat{\beta})\right)\left[1-\left(1-\nu(\bar{c}\hat{\beta})\right)\bar{c}\kappa\right]\end{eqnarray}
where $\nu(x)=(\exp(-x)-1+x)/x$. Substituting for $\nu(x)$, we have
\begin{equation}
\P(\text{success})\leq P_{p}\left(\frac{\lambda (1-\exp(-\bar{c}\hat{\beta}))}{\bar{c}\hat{\beta}} \right)\left[1-\left(1-\exp(-\bar{c}\hat{\beta})\right)\frac{\kappa}{\hat{\beta}}\right]\label{eq:tight-upper}\end{equation}
\eqref{eq:tight-upper} is a tighter bound than the bound in Lemma
\ref{lem:[Upper-bound] }, but not easily computable due to the presence
of $\kappa$ (for a given $R,T$ and $\sigma$, $\kappa$ and $\beta^{*}$
are constants). In \eqref{eq:tight-upper}, the outage due to the
interference by the transmitting cluster is also taken into account.

The proof of Lemmata \ref{lem:lower} and \ref{lem:[Upper-bound] }
also indicates that it is only by conditioning on the event that there
is a point at the origin that the success probability of Neyman-Scott
cluster processes can be lower than the Poisson process of the same
intensity. This implies that the cluster around the transmitter causes
the maximum {}``damage''. So as the receiver moves away from the
transmitter, the Neyman-Scott cluster process has a better success
probability than the PPP. So, it  is not true in general that cluster
processes have a lower success probability than PPPs of the same intensity.
\begin{figure}[h]
 
 \centering \includegraphics[scale=0.5]{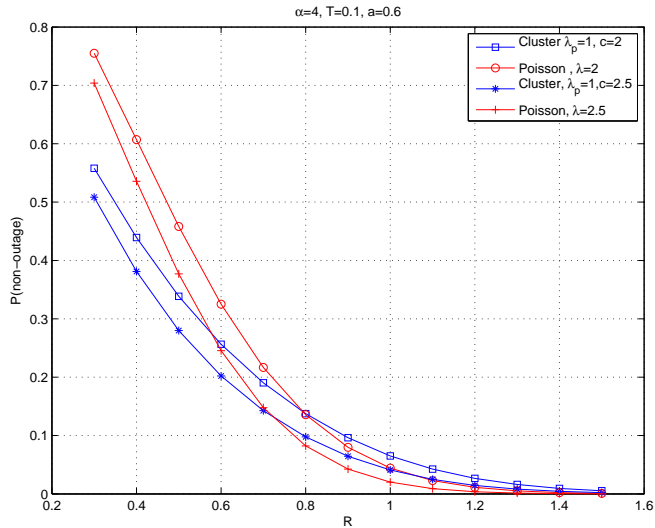}
 \caption{Comparison of  success probability for cluster and Poisson process of intensity $2$}
\label{fig:comp1}  

\end{figure}
For example from Figure~\ref{fig:comp1}, we see that for $R<0.8$,
the PPP has a better success probability than the Matern process.
 In Subsection \ref{sub:cross-over} we give a more detailed analysis,
which reveals that a PPP with intensity $\lambda_{p}\bar{c}$ has
a lower success probability than  a clustered process of the same
intensity {\em for large transmit-receiver distances.} On the other
hand, for small $R$, the success probability of the PPP is higher.

\subsection{\label{sub:cross-over} Clustering Gain $G(R)$}

In this subsection we compare the performances of a clustered network
and a Poisson network of the same intensity with Rayleigh fading. We deduce how the clustering
gain  depends on the transmitter receiver distance. We use the following notation,
\begin{eqnarray}
P_{1}(R,\bar{c},\lambda_{p}) & \stackrel{\Delta}{=} & \exp\Big(-\lambda_{p} \int_{\R^{2}}1-\exp(-\bar{c}\beta(R,y)) \d y\Big)\\
P_{2}(R,\bar{c}) & \stackrel{\Delta}{=} & \int_{\R^{2}}\exp(-\bar{c}\beta(R,y))f(y)\d y\end{eqnarray}
 So $\mathbb{P}(\text{success})=P_{1}(R,\bar{c},\lambda_{p})P_{2}(R,\bar{c})$.
$P_{2}$ is the probability of success due to the presence of the
cluster at the origin near the transmitter. $P_{1}$ is the probability
of success in the presence of other clusters. Interference from these
other clusters contributes more to the outage when $R$ is large.
This is also intuitive, since as the receiver moves away from the
transmitting cluster, the interference from the other clusters starts
to dominate. We define the {\em clustering gain} $G(R)$ as \[
G(R)=\frac{P_{1}(R,\bar{c},\lambda_{p})P_{2}(R,\bar{c})}{P_{p}(\lambda_{p}\bar{c})}\]
The   fluctuation  of $G(R)$ around unity indicates  the existence of a crossover
point $R^{*}$ below which the PPP performs better than clustered
process and vice versa. The values of $G(R)$ at the  origin and infinity
indicate the gain of scheduling transmitters as clusters instead of
being spread uniformly on  the plane. So it is beneficial to induce
logical clustering of transmitters by MAC if $G(R)>1$.

We first consider $G(R)$ for large $R$, \ie, $\lim_{R\rightarrow\infty}G(R)$.
By the dominated convergence theorem and \eqref{eq:assume2}, we have
\begin{eqnarray}
\lim_{R\rightarrow\infty}P_{2}(R,\bar{c}) & = & \int_{\R^{2}}\exp\Big(-\bar{c}\int_{\R^{2}}\lim_{R\rightarrow\infty}\frac{f(x)}{1+\frac{g(R)}{Tg(x-y-R)}}\d x\Big)f(y)\d y\nonumber \\
 & = & \exp\Big(\frac{-\bar{c}}{1+1/T}\Big)\label{eq:claim1}\end{eqnarray}
 Also from the derivation of upper bound we have $P_{1}(R,\bar{c},\lambda_{p})\leq P_{p}\left(\frac{\lambda}{1+\bar{c}\hat{\beta}}\right)$.
Hence from the definition of $P_{p}(x)$ we have, $\lim_{R\rightarrow\infty}P_{1}(R,\bar{c},\lambda_{p})=0$.
Hence for large $R$,
\begin{equation}
P_{1}(R,\bar{c},\lambda_{p})<P_{2}(R,\bar{c})\label{eq:231}\end{equation}
So for large $R$, most of the damage is done by transmitting nodes
other than the cluster in which the intended transmitter lies. 
\begin{lemma}
\label{lem:limit}\begin{eqnarray}
\lim_{R\rightarrow\infty}\frac{P_{p}(\lambda_{p}\bar{c})}{P_{1}(R,\bar{c},\lambda_{p})} & = & 0\end{eqnarray}
\end{lemma}
\begin{proof}
See Appendix \ref{sub:Proof-of-Lemma-limit}
\end{proof}
Hence for large $R$, $\frac{P_{p}(\lambda_{p}\bar{c})}{P_{1}(R,\bar{c},\lambda_{p})}\leq\exp(\frac{-\bar{c}}{1+1/T})$.
From (\ref{eq:claim1}) we have $P_{p}(\lambda_{p}\bar{c})\leq P_{1}(R,\bar{c},\lambda_{p})P_{2}(R,\bar{c})$,
for large $R$, \ie, $G(R)>1$ for large transmit-receive distance.
We have $\lim_{R\rightarrow\infty}G(R)=\infty$. Hence the\emph{ Poisson
point process with intensity $\lambda_{p}\bar{c}$, has a lower success
probability than the clustered process of the same intensity for
large transmit receiver distances.}

For  small $R$, \emph{$G(R)$ }depends on the behavior of the path
loss function, $g(x)$ at $\Vert x\Vert=0$. We consider the two cases
when the channel function is singular at the origin or not.

\subsubsection{$\lim_{\Vert x\Vert\rightarrow0}g(x)=\infty$}

In this case we observe that $G(0)=1$. But at small $R$, $G(R)$
is less than $1$. We have the following lemma. 

\begin{lemma}
\label{lemma4} If $(f*f)(x)>\Vert x\Vert$ for {\em small} $\Vert x\Vert$
and $g(x)=\Vert x\Vert^{-\alpha}$, then for {\em small} $R$,
\begin{eqnarray}
\mathbb{\mathbb{P}}(\text{success}) & \leq & P_{p}(\lambda_{p}\bar{c})\end{eqnarray}

\end{lemma}
\begin{proof}
See Appendix \ref{sub:Proof-of-Lemma4}.  
\end{proof}
Note that  $f(x)$ for Matern and Thomas cluster process have the required property.
Hence when $g(x)=\Vert x\Vert^{-\alpha}$,\emph{ the PPP with intensity
$\lambda_{p}\bar{c}$, has a higher success probability than the clustered
process of the same intensity for small transmit receiver distance}.
Lemma \ref{lemma4} and the fact that $G(\infty)=\infty$ also indicate
the existence of a crossover point $R^{*}$ between the success curves
of the PPP and the cluster process. So it is not true in general that
the performance of the clustered process is better or worse than that
of the Poisson process. This is because, for the same intensity, a
clustered process will have clusters of transmitters (where interference
is high) and also vacant areas (where there are no transmitters and
interference is low), whereas in a Poisson process, the transmitters
are uniformly spread.

\subsubsection{$\lim_{\Vert x\Vert\rightarrow0}g(x)=\hat{g}<\infty$}

\begin{figure} 
\begin{centering}
\includegraphics[scale=0.6]{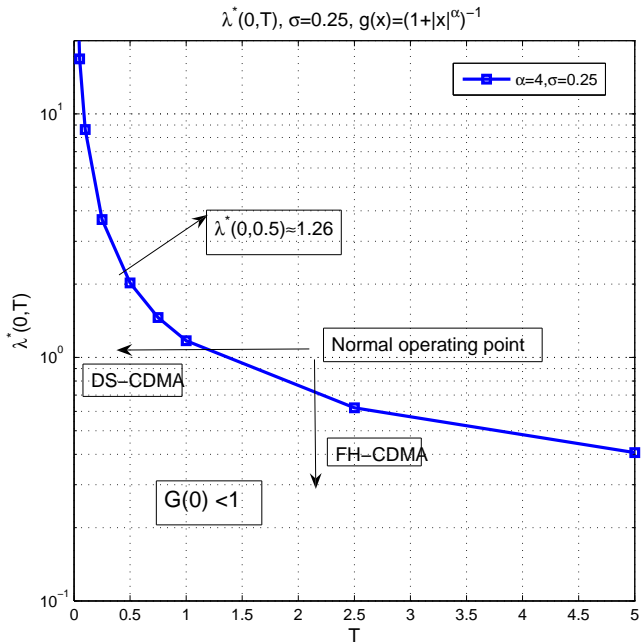}
\par\end{centering}
\caption{$\lambda^{*}(0,T)$ versus $T$ for $g(x)=(1+\Vert x\Vert^{4})^{-1}$,
$\sigma=0.25$. The region below the curve  consists of all the pairs of $(T,\lambda=\lambda_p\bar{c})$ such that $G(0)<1$.   ``Normal operating point'' denotes a  pair  $(T,\lambda)$ that  lies above the curve $(T,\lambda^{*}(0,T))$. Suppose we use FH-CDMA, the total intensity decreases by a factor of spreading gain and  hence we move vertically downwards into the $G(0)<1$ region. If DS-CDMA is used, the threshold $T$ decreases by  a factor of spreading gain and hence we move horizontally towards the left   into the $G(0)<1$ region.}
\label{fig:threshold}
\end{figure}

\begin{figure} 
\begin{centering}
\includegraphics[scale=0.6]{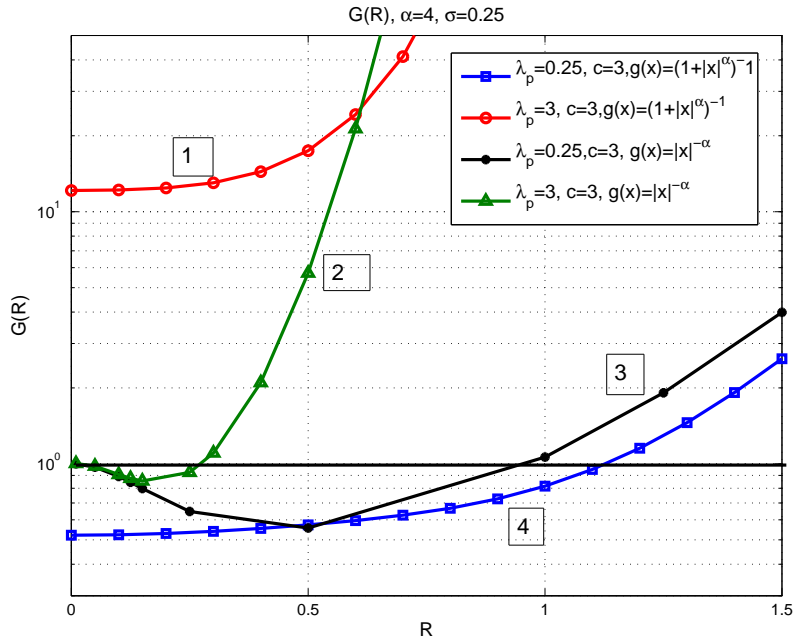}
\par\end{centering}
\caption{$G(R)$ versus $R$, $\alpha=4,\sigma=0.25$. Observe that the  gain curves $\#2$ and $\#3$, which correspond to the singular channel, start at $1$ decrease and then increase above unity. For the gain curve $\#4$,  the total intensity of transmitters is $3*0.25 =0.75$ which is less than the  threshold $\lambda^{*}(0,0.5)\approx1.26$. Hence the gain curve  for this case starts below  unity at $R=0$  and then increases. For the gain curve $\#1$ the total intensity is $9>1.26$.  By chance,  in the present case  the gain curve $\#1$ starts  around $10$ and increases.   }
\label{fig:G_R}
\end{figure}

$P_{1}(R,\bar{c},\lambda_{p})$ can be written as \[
P_{1}(R,\bar{c},\lambda_{p})=P_{p}(\lambda_{p}\bar{c})\exp\Big(\lambda_{p}\int_{\R^{2}}\underbrace{\sum_{n=2}^{\infty}\frac{(-1)^{n}}{n!}\bar{c}^{n}\beta(R,y)^{n}}_{>0}\d y\Big)\]
Hence $G(R)$ can also be written as follows\begin{eqnarray}
G(R) & = & P_{2}(R,\bar{c})\exp\left(\lambda_{p}\bar{c}\int_{\R^{2}}\beta(R,y)\eta(\bar{c},R,y)\d y\right)\label{eq:onnne}\end{eqnarray}
where $\eta(\bar{c},R,y)=\nu(\bar{c}\beta(R,y))$, with $\nu(x)=(\exp(-x)-1+x)/x$.
Observe that $0\leq\eta(\bar{c},R,y)\leq1,\forall x>0$. If the total
density of the transmitters is fixed \ie, $\lambda=\lambda_{p}\bar{c}$
is constant, how does $G(R)$ behave with respect to $\bar{c}$? We
have the following lemma which characterizes the monotonicity of $G(R)$
with respect to $\bar{c}$.
\begin{lemma}
\label{lem:GR}Given $\lambda=\lambda_{p}\bar{c}$ is constant,
$G(R)$ is decreasing with $\bar{c}$, \ie, $\frac{\d G(R)}{\d \bar{c}}\leq0\ ,\ \forall\bar{c}>0$
iff $\lambda\leq\lambda^{*}(R,T)$, where \[
\lambda^{*}(R,T)=\frac{2\int_{\R^{2}}\beta(R,y)f(y)\d y}{\int_{\R^{2}}\beta(R,y)^{2}\d y}\]
\end{lemma}
\begin{proof}
From \foreignlanguage{english}{\eqref{eq:onnne},} \foreignlanguage{english}{\begin{eqnarray}
G(R) & = & P_{2}(R,\bar{c})\exp\left[\lambda_{p}\bar{c}\int_{\R^{2}}\beta(R,a)\eta(\bar{c},R,a)\d a\right]\nonumber\\
 & = & \int_{\R^{2}}\exp\Big(-\bar{c}\beta(R,y)+\lambda\int_{\R^{2}}\beta(R,a)\eta(\bar{c},R,a)\d a\Big)f(y)\d y\end{eqnarray}
 We have $\frac{\d \eta(\bar{c},R,z)}{\d \bar{c}}|_{\bar{c}=0}=\beta(R,z)/2$
and $\frac{\d \eta(\bar{c},R,z)}{\d \bar{c}}$ is decreasing in $\bar{c}$.
\begin{eqnarray}
\frac{\d G(R)}{\d 	\bar{c}} & = & \int_{\R^{2}}\left[-\beta(R,y)+\lambda\int_{\R^{2}}\beta(R,a)\frac{\d \eta(\bar{c},R,a)}{\d \bar{c}}\d a\right]\exp\Big(-\bar{c}\beta(R,y)+\lambda\int_{\R^{2}}\beta(R,a)\eta(\bar{c},R,a)\d a\Big)f(y)\d y\nonumber\\
 & = & \exp[\lambda\int_{\R^{2}}\beta(R,a)\eta(\bar{c},R,a)\d a]\underbrace{\int_{\R^{2}}\Big[-\beta(R,y)+\lambda\int_{\R^{2}}\beta(R,z)\frac{\d \eta(\bar{c},R,a)}{\d \bar{c}}\d a\Big]\exp\Big(-\bar{c}\beta(R,y)\Big)f(y)\d y}_{T_{2}(\bar{c})}\nonumber\end{eqnarray}
 Since $\eta'(\bar{c},R,z)$ is decreasing in $\bar{c}$, we have
$T_{2}(\bar{c})$ is decreasing in $\bar{c}$. So a necessary and
sufficient condition for $\frac{dG(R)}{d\bar{c}}\leq0\ \forall\bar{c}>0$
is $T_{2}(0)\leq0$. We want \begin{eqnarray}
T_{2}(0) & = & \int_{\R^{2}}\left[-\beta(R,y)+\frac{\lambda}{2}\int_{\R^{2}}\beta^{2}(R,z)\d z\right]f(y)\d y\leq0\nonumber\\
 & \Rightarrow & \lambda\leq\frac{2\int_{\R^{2}}\beta(R,y)f(y)\d y}{\int_{\R^{2}}\beta^{2}(R,z)\d z}\end{eqnarray}
}
\end{proof}
Remarks:
\begin{enumerate}
\item Since\emph{ $\beta(0,y)\neq0$, }we have that, $G(0)$ is increasing
with $\lambda_{p}$ (like $\ \exp(\lambda_{p}))$, and hence will
be greater than $1$ at some $\lambda_{p}$ for a fixed $\bar{c}$.
\item We have $\lim_{\bar{c}\rightarrow0}G(R)=1$ and specifically $G(0)=1$
at $\bar{c}=0$.
\item From Lemma \ref{lem:GR} and Remark $2$ we can deduce $G(R)<1,\ \forall\bar{c}>0$
if $\lambda<\lambda^{*}(R,T)$ \ie, the gain $G(R)$ decreases
from $1$ with increasing $\bar{c}$ if the total intensity of transmitters
is less than $\lambda^{*}(R,T)$.
\item Since $G(R)$ is continuous with respect to $R$, $G(R)$ is close
to $G(0)$ for small $R$.
\item From \foreignlanguage{american}{Figure \ref{fig:G_R}}, we observe
that $G(R)$ increases monotonically with $R$. 
\end{enumerate}
In Figure \ref{fig:threshold}, $\lambda^{*}(0,T)$ is plotted against
$T$.\foreignlanguage{american}{ }We provide some heuristics as to
when logical clustering does not perform better than  a uniform distribution
of points: 
\begin{itemize}
\item The exact value of $R$ at which $G(R)$ crosses $1$  is difficult
to find analytically due to the highly nonlinear nature of $G(R)$.
If such a crossover point exists (depends on the path-loss model) we
will denote it by $R^{*}$.
 
\item If $g(x)=\Vert x\Vert^{-\alpha}$, it is better to induce logical
clustering by the MAC scheme if the link distance is larger than
$R^{*}$. Otherwise it is better to schedule the transmissions so
that they are scattered uniformly on the plane.
 
\item If $g(0)<\infty$  and for a constant intensity
$\lambda_{p}\bar{c}$, it is always beneficial to induce clustering
for long-hop transmissions. When $R$ is
small the answer depends on the total intensity  $\lambda_{p}\bar{c}$.
If  $\lambda_{p}\bar{c}<\lambda^{*}(0,T)$
then $G(0)<1$ by observation $3$, and hence $G(R)<1$ for \emph{small}
$R$ by observation $4$. Also when $\lambda_{p}\bar{c}<\lambda^{*}(0,T)$,
it is better to reduce logical clustering by decreasing $\bar{c}$
and increasing $\lambda_{p}$, since $G(0)$ 
 is a decreasing function of $\bar{c}$.  From Figure \ref{fig:threshold} 
we observe that  $\lambda^{*}(0,0.5)\approx1.26$ when $g(x)=(1+\Vert x\Vert^{4})^{-1}$
and $\sigma=0.25$. In Figure \ref{fig:G_R}, $G(R)$ is plotted for
$\lambda_{p}\bar{c}=0.75,9$ for the same values of $\sigma,\alpha$
and the same channel function as of Figure~\ref{fig:threshold}. When $\lambda_{p}\bar{c}=9>\lambda^{*}(0,0.5)$, we observe
that the gain curve $G(R)$ is approximately $10$ at the origin and
increases. When $\lambda_{p}\bar{c}=0.75<\lambda^{*}(0,0.5)$, $G(R)$ starts around
$0.25$ and crosses $1$ at $R\approx1.2$. We also observe that $G(R)$,
for the non-singular $g(x)$, seem to increase monotonically. We also
observe that the gain function for $g(x)=\Vert x\Vert^{-\alpha}$
decreases from $1$ initially and then increases to infinity.
 
\item For DS-CDMA, the value of $T$ is smaller by a factor equal to the  spreading
gain. From Figure \ref{fig:threshold},
we observe that the threshold $\lambda^{*}(0,T)$ for clustering
to be beneficial at small distances increases with decreasing $T$.
Hence for a constant intensity of transmissions
$\lambda_{p}\bar{c}$, the benefit of clustering decreases with increasing
spreading gain for small link distances. So for DS-CDMA (for a large
spreading gain) it is better to make the transmissions uniform on
the plane for smaller link distances and cluster the transmitters
for long-range communication. 
\item For FH-CDMA,  the
total number of transmissions $\lambda_{p}\bar{c}$ is reduced by
the spreading gain while $T$ remains constant (see Figure \ref{fig:threshold}).
Hence $\lambda_{p}\bar{c}<\lambda^{*}(0,T)$  for small distances
and  one can draw similar conclusions  as
that of DS-CDMA. The relative gain between FH-CDMA and DS-CDMA  with
clustering  is more difficult to characterize analytically. 
\end{itemize}

\section{Transmission Capacity  of Clustered Transmitters \label{sec:Transmission-Capacity}}

It is important to understand the performance of ad hoc wireless networks.
Transmission capacity was introduced in \foreignlanguage{english}{\cite{weber:2005,WebAnd2006d,WebAnd2006a}
and  is defined as the product of the maximum density of successful
transmissions and  their data rate, given an outage constraint. More
formally, if the intensity of the contending transmitters is $\lambda$
with an outage threshold $T$ and a bit rate $b$ bits per second
per hertz, then the transmission capacity at a fixed distance $R$
is given by\begin{equation}
C(\epsilon,T)=b(1-\epsilon)\sup_{\lambda}\{\lambda:P(\lambda,T)\geq1-\epsilon\}\label{eq:tx-capacity}\end{equation}
 where $P(\lambda,T)$ denotes the success probability of a given
transmitter receiver pair. More discussion about the transmission
capacity and its relation to other metrics like transport capacity
is provided in \cite{WebAnd2006d}.  Note that  the results proved in
\cite{weber:2005,WebAnd2006d,WebAnd2006a} are for Poisson arrangement
of transmitters. }

In this section we evaluate the transmission capacity when the transmitters
are arranged as a Poisson clustered process. We prove that for small
values of $\epsilon$, the transmission capacity of the clustered
process coincides with that of the Poisson arrangement of nodes. We
also show that   care should be taken in defining transmission
capacity for general distribution of nodes. For notational convenience
we shall assume $b=1$. 
For the clustered process, $P(\lambda,T)$ denotes the success probability
of the  cluster process with intensity $\lambda=\lambda_{p}\bar{c}$ and
threshold $T$. Let $P_{l}(\lambda,T),\: P_{u}(\lambda,T)$ denote
lower and upper bounds of the success probability $P(\lambda,T)$
and the corresponding sets $A_{l},A_{u}$ defined by $A_{\chi}:=\{\lambda:P_{\chi}(\lambda,T)\geq1-\epsilon\}$
for $\chi\in\{l,u\}$. We then have $A_{l}\subset A\subset A_{u}$
which implies  \begin{equation}
\sup A_{l}\leq\sup A\leq\sup A_{u}\label{eq:imp-tx}.\end{equation}
Let $C_{l}(\epsilon,T)=\sup A_{l}$ and $C_{u}(\epsilon,T)=\sup A_{u}$
denote lower and upper bounds to the transmission capacity.

For a PPP we have from \eqref{eq:poisson}, $P_{p}(\lambda,T)=\exp(-\lambda\beta_{I})$
($\beta_{I}$ does not depend on $\lambda$). Hence the transmission
capacity of a PPP denoted by $C_{p}(\epsilon,T)$ is given by \begin{eqnarray}
C_{p}(\epsilon,T) & = & \frac{1-\epsilon}{\beta_{I}}\ln\left(\frac{1}{1-\epsilon}\right)\nonumber\\
 & \approx & \frac{\epsilon(1-\epsilon)}{\beta_{I}},\ \epsilon\ll1\end{eqnarray}
For Neyman-Scott cluster processes, the intensity $\lambda=\lambda_{p}\bar{c}$.
We first to try to consider both $\lambda_{p}$ and $\bar{c}$ as
optimization parameters for the transmission capacity, i.e. \begin{eqnarray}
C(\epsilon,T) & := & (1-\epsilon)\sup\{\lambda_{p}\bar{c}:\lambda_{p}>0,\bar{c}>0,\text{outage-constraint}\}\end{eqnarray}
 without individually constraining the parent node density or the
average number of nodes per cluster.
\begin{lemma}
\label{lem:lower-cap}The transmission capacity of Poisson clustered
processes is lower bounded by the transmission capacity of the  PPP, \begin{equation}
C(\epsilon,T)\geq C_{l}(\epsilon,T)=C_{p}(\epsilon,T)\label{eq:lower-cap}\end{equation}
 \end{lemma}
\begin{proof}
From Lemma \ref{lem:lower}, we have $P_{l}(\lambda,T)=P_{p}(\lambda_{p}\bar{c})P_{p}(\bar{c}\hat{f}^{*})$.
So to get a lower bound, from (\ref{eq:imp-tx}) we have to find  
  \begin{eqnarray}
\sup\Big\{\lambda_{p}\bar{c}: \lambda_{p}\bar{c}+\bar{c}\hat{f}^{*} & \leq & \frac{1}{\beta_{I}}\ln\left(\frac{1}{1-\epsilon}\right)=\frac{C_{p}(\epsilon,T)}{1-\epsilon} \Big\}
\end{eqnarray}
 This maximum value of $\lambda_{p}\bar{c}$ is attained when, $\lambda_{p}\rightarrow\infty$,
while $\bar{c}\rightarrow0$, such that $\bar{c}\lambda_{p}=C_{p}(\epsilon,T)(1-\epsilon)^{-1}$.
So we have $C_{l}(\epsilon,T)=C_{p}(\epsilon,T)$.
\end{proof}
Also observe that   $\lambda_{p}\rightarrow\infty$ and
$\bar{c}\rightarrow0$. This corresponds to the scenario in which
the clustered process degenerated to a PPP. We also have the following
upper bound.\foreignlanguage{american}{ }

\begin{lemma}
\label{lem:upper-cap}Let $\rho(T)=k/\hat{\beta}$ with $k=\int\beta(R,y)f(y)\d y$.
For $\epsilon<1-e^{-\rho(T)}$, we have \begin{eqnarray}
C(\epsilon,T) & \leq & C_{u}(\epsilon,T)=C_{p}(\epsilon,T)\end{eqnarray}

\end{lemma}
\begin{proof}
See Appendix \ref{sub:Proof-upper-cap}.
\end{proof}
\begin{thm}
For $\epsilon\leq1-e^{-\rho(T)}$ we have $C(\epsilon,T)=C_{p}(\epsilon,T)$. 
\end{thm}
\begin{proof}
Follows from the Lemmata~\ref{lem:lower-cap} and \ref{lem:upper-cap}.
\end{proof}
From the above two proofs, when $\epsilon$ is small, the transmission
capacity is equal to the Poisson process of same intensity. This capacity
is  achieved when $\lambda_{p}\rightarrow\infty$ and $\bar{c}\rightarrow0$.
This is the scenario in which the cluster process becomes a  
PPP. This is due to the definition of the transmission capacity
as $C(\epsilon,T):=\sup\{\lambda_{p}\bar{c}:\lambda_{p}>0,\bar{c}>0,\text{outage-constraint}\}$
where we have two variables to optimize over. 

Instead we may fix $\lambda_{p}$ as constant and find the transmission
capacity with respect to $\bar{c}$. So  we define constrained transmission
capacity as\begin{eqnarray}
C^{*}(\epsilon,T):=\lambda_{p}(1-\epsilon)\sup\{\bar{c}:\bar{c}>0,\text{outage-constraint}\}\end{eqnarray}
 We have the following bounds for $C^{*}(\epsilon,T)$

\begin{thm}
\begin{eqnarray}
\frac{\lambda_{p}C_{p}(\epsilon,T)}{\lambda_{p}+\hat{f}^{*}} & \leq C^{*}(\epsilon,T)\leq & \frac{\lambda_{p}C_{p}(\epsilon,T)}{\max\left\{ 0,\lambda_{p}-\frac{\hat{\beta}}{\beta_{I}}\ln\left(\frac{1}{1-\epsilon}\right)\right\} }\end{eqnarray}

\end{thm}
\begin{proof}
From the lower bound on $\P(\text{success})$, we have to find  
  \begin{eqnarray}
\sup\Big \{\bar{c}:\lambda_{p}\bar{c}+\bar{c}\hat{f}^{*} & \leq & \frac{1}{\beta_{I}}\ln\left(\frac{1}{1-\epsilon}\right)=\frac{C_{p}(\epsilon,T)}{1-\epsilon}\Big \}\end{eqnarray}
So we have $C_{l}^{*}(\epsilon,T)=C_{p}(\epsilon,T)/(\hat{f}^{*}+\lambda_{p})$.

From the upper bound on $P(\text{success})$,we have to find   \begin{eqnarray}
\sup\Big \{\bar{c}:\frac{\lambda_{p}\bar{c}}{1+\bar{c}\hat{\beta}} & \leq & \frac{1}{\beta_{I}}\ln\left(\frac{1}{1-\epsilon}\right)=\frac{C_{p}(\epsilon,T)}{1-\epsilon}\Big \}\end{eqnarray}
 \end{proof}

\begin{figure*}
\begin{centering}
\includegraphics[scale=0.5]{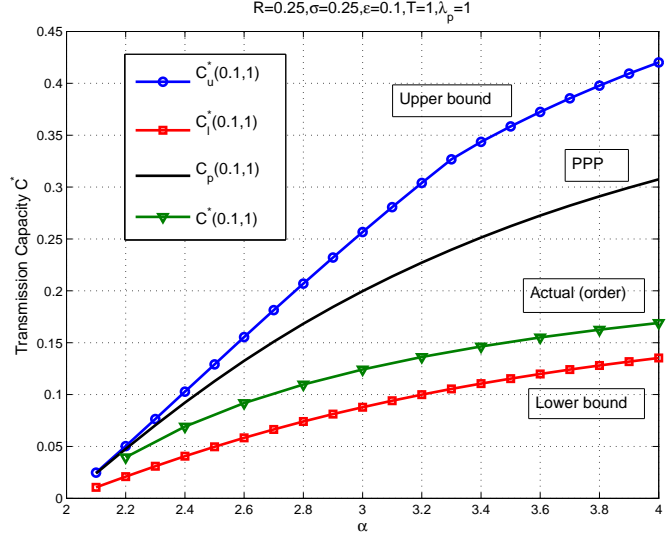}\caption{Upper and lower bounds of $C^{*}(\epsilon,T)$ versus $\alpha$, $g(x)=\Vert x\Vert^{-\alpha},T=1,\sigma=0.25,\epsilon=0.1,\lambda_{p}=1$}
\label{fig:TXC_comp1} 
\par\end{centering}
\end{figure*}
 One can also derive an order approximation to the constrained transmission
capacity when $\epsilon$ is very small. We have the following order
 approximation to transmission capacity.

\begin{prop}
When $\lambda_{p}$ is fixed, the constrained transmission capacity
is given by \begin{eqnarray}
C^{*}(\epsilon,T) & = & (1-\epsilon)\left(\frac{\epsilon\lambda_{p}}{\lambda_{p}\beta_{I}+k}+o(\epsilon )\right)\end{eqnarray}
when $\epsilon\rightarrow0$.
\end{prop}
\begin{proof}
Let $\gamma(\bar{c)}$ denote the outage probability, \ie,\begin{eqnarray}
\gamma(\bar{c}) & = & 1-\exp\Big\{-\lambda_{p}\int_{\R^{2}}1-\exp[-\bar{c}\beta(R,y)]\d y\Big\}\int_{\R^{2}}\exp(-\bar{c}\beta(R,y))f(y)\d y\end{eqnarray}
We have $\d \gamma(\bar{c})/\d \bar{c}>0$, which implies $\gamma(\bar{c})$
is increasing and invertible and hence $C^{*}(\epsilon,T)=\lambda_{p}(1-\epsilon)\gamma^{-1}(\epsilon)$.
We approximate $\gamma^{-1}(\epsilon)$ for small $\epsilon$ by the Lagrange
inversion theorem. Observe that $\gamma(\bar{c})$ is a smooth function
of $\bar{c}$ and all derivatives exist. Expanding $\gamma^{-1}(\epsilon)$
around $\epsilon=0$ by the  Lagrange inversion theorem and using $\gamma(0)=0$ yields \begin{eqnarray}
\gamma^{-1}(\epsilon) & = & \sum_{n=1}^{\infty}\frac{d^{n-1}}{d\bar{c}^{n-1}}\left(\frac{\bar{c}}{\gamma(\bar{c})}\right)^{n}\left|_{\bar{c}=0}\frac{\epsilon^{n}}{n!}\right.\label{eq:444}\\
 & = & \frac{\bar{c}\epsilon}{\gamma(\bar{c})}\left|_{\bar{c}=0}\right.+o(\epsilon)\nonumber \\
 & \stackrel{(a)}{=} & \frac{\epsilon}{\lambda_{p}\beta_{I}+k}+o(\epsilon)\nonumber \end{eqnarray}
where $(a)$ follows by applying  de L'H\^opital's rule. %
{}
\end{proof}
We have the following observations
\begin{enumerate}
\item The constrained transmission capacity increases (slowly) with $\lambda_{p}$. 
\item We also observe that the constrained transmission capacity for the
cluster process is always less than that of  a Poisson network (see Figure
\ref{fig:TXC_comp1}) and approaches $C_{p}(\epsilon,T)$ as $\lambda_{p}\rightarrow\infty$. 
\item When FH-CDMA with intra-cluster frequency hopping is utilized, we
have the cluster intensity $\bar{c}$ reduced by a factor $M$ (spreading
gain). One can easily obtain the constrained transmission capacity
of this system to be \[
C_{FH}^{*}(\epsilon,T)=(1-\epsilon)\left(\frac{\epsilon\lambda_{p}M}{\lambda_{p}\beta_{I}+k}+o(\epsilon)\right)\]
When DS-CDMA is used, the constrained transmission capacity is $C_{DS}^{*}(\epsilon,T)=C^{*}(\epsilon,T/M)$.
When the transmitters are spread as a Poisson point process, we have
from \cite{haenggi-com-2007,weber2005tcw} \[
\ln\left(\frac{C_{FH}(\epsilon,T)}{C_{DS}(\epsilon,T)}\right)=(1-2/\alpha)\ln(M).\]
 In Figure \ref{fig:gain_fh_ds}, we plot $\ln(C_{FH}^{*}(\epsilon,T)/C_{DS}^{*}(\epsilon,T))/\ln(M)$
with respect to spreading gain $M$, when the path loss function is
 $g(x)=\Vert x\Vert^{-\alpha}$ and $\epsilon=0.01$. From the figure,
we observe a similar $M^{1-2/\alpha}$ gain, even in the case of clustered
transmitters. %
\begin{figure}[h]
\begin{centering}
\includegraphics[scale=0.5]{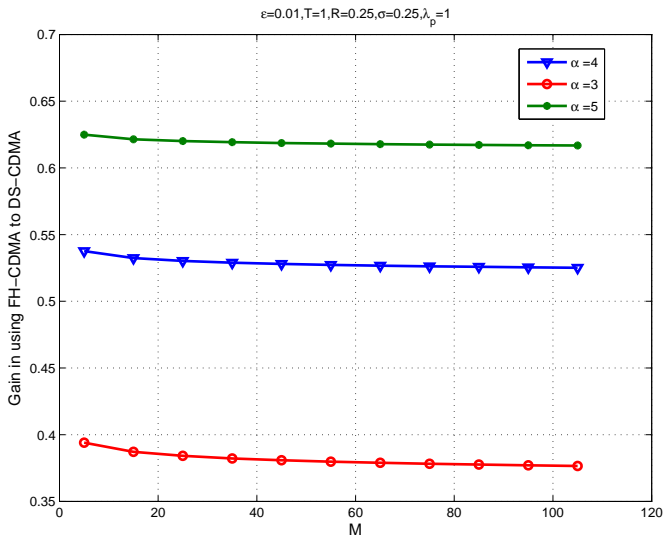}
\par\end{centering}
\caption{$\ln(C_{FH}^{*}(\epsilon,T)/C_{DS}^{*}(\epsilon,T))/\ln(M)$ versus
$M$ for $\epsilon=0.01,\lambda_{p}=1$}
\label{fig:gain_fh_ds}
\end{figure}
\end{enumerate}

\section{\label{sec:Conclusions}Conclusions}

Previous work characterizing interference, outage, and transmission 
capacity
in large random networks exclusively focused on the homogeneous Poisson 
point
process as the underlying node distribution. In this paper, we extend 
these results
to clustered processes. Clustering may be geographical, \ie, given by 
the spatial
distribution of the nodes, or it may be induced logically by the MAC 
scheme.
We use tools from stochastic geometry and Palm probabilities to obtain
the conditional Laplace transform of the interference.   Upper and lower bounds
are obtained for the CCDF of the interference, for any stationary distribution of nodes 
and fading.  We have shown that the  distribution of interference depends 
heavily on the  path-loss model considered.  In particular, the existence of a singularity in the model greatly
affects the results.
 This conditional Laplace transform is then used to obtain the  probability of success in
a clustered network with Rayleigh fading.   We show   clustering  the transmitters is 
always beneficial for large link distances, while the clustering gain at smaller link distances
depends  on the path-loss model.
 The transmission capacity of clustered networks is equal to the one 
for homogeneous
networks. However, care must be taken when defining this capacity since 
clustered
processes have two parameters to optimize over. We
also show that the transmission capacity of clustered network is equal
to the Poisson distribution of nodes. 
We anticipate that the analytical techniques used in this work will be 
useful for other
problems as well. In particular the conditional generating functionals 
are likely to
find wide applicability.
\section*{Acknowledgments}
The support of the NSF (grants CNS 04-47869, CCF 
05-15012, and DMS 505624) and the
DARPA/IT-MANET program (grant W911NF-07-1-0028) is gratefully 
acknowledged.

\bibliographystyle{IEEEtran}
\bibliography{point_process}

\begin{appendix}

\subsection{\label{sub:Proof-of-Lemma-scaling}Proof of Lemma \ref{lem:scaling}:}

\begin{proof}
We first evaluate the asymptotic behavior  of $\mathcal{G}\left(F_{h }\left(\frac{y}{g(.-z)}\right)\right)$.
Let $v(x)=F_{h }\left(\frac{y}{g(x-z)}\right)$. We have \begin{eqnarray}
M\left(\int_{\R^{2}}v(x+u)f(u)\d u\right) & = & \exp\left(-\bar{c}\int_{\R^{2}}[1-F_{h }\left(\frac{y}{g(x+u-z)}\right)]f(u)\d u\right)\nonumber \\
 & \stackrel{(a)}{\sim} & 1-\bar{c}\int_{\R^{2}}\left[1-F_{h }\left(\frac{y}{g(x+u-z)}\right)\right]f(u)\d u\label{eq:1234}\end{eqnarray}
where $(a)$ follows from the fact that $\ \exp(-x)=1-x+\mathcal{O}(x^{2})$
for $x$ close to $0$ and $(1-F_{h })\rightarrow0$ for large
$y$. By a similar expansion of $\exp$, \eqref{eq:1234} and the dominated convergence
theorem, we have \begin{eqnarray}
 &  & \exp\left(-\lambda_{p}\int_{\R^{2}}1-M(\int_{\R^{2}}v(x+u)f(u)\d u)\d x\right)\nonumber\\
 & \sim & 1-\lambda_{p}\bar{c}\int_{\R^{2}}\int_{\R^{2}}[1-F_{h }\left(\frac{y}{g(x+u-z)}\right)]f(u)\d u\d x\nonumber\\
 & = & 1-y^{-2/\alpha}\lambda_{p}\bar{c}\int_{\R^{2}}[1-F_{h }(\Vert x\Vert^{\alpha})]\d
 x \label{12345}
\end{eqnarray}
By change of variables and using $\lim_{y\rightarrow\infty}[1-F_{h }(y)]y^{2/\alpha}=0$
\cite[p.198]{folland}, we have \[
\int_{\R^{2}}[1-F_{h }(\Vert x\Vert^{\alpha})]\d x=\pi\int_{0}^{\infty}\nu^{2/\alpha}\d 	F_{h }(\nu)\]
 We similarly have \begin{eqnarray}
 &  & \int_{\R^{2}}M\left(\int_{\R^{2}}v(x+u)f(u)\d u\right)f(x)\d x\nonumber\\
 & \sim & 1-\bar{c}\int_{\R^{2}}\left(1-\int_{\R^{2}}F_{h }\left(\frac{y}{g(x+u-z)}\right)f(u)\d u\right)f(x)\d x\nonumber\\
 & = & 1-\bar{c}\int_{\R^{2}}\int_{\R^{2}}\left[1-F_{h }\left(y\Vert x+u-z\Vert^{\alpha}\right)\right]f(u)f(x)\d u\d x\nonumber\\
 & = & 1-\bar{c}\int_{\R^{2}}\left[1-F_{h }\left(y\Vert x\Vert^{\alpha}\right)\right](f*f)(x+z)\d x\nonumber\\
 & = & 1-\bar{c}y^{-2/\alpha}\int_{\R^{2}}\left[1-F_{h }\left(\Vert x\Vert^{\alpha}\right)\right](f*f)\left(\frac{x}{y^{1/\alpha}}+z\right)\d x\nonumber\\
 & \stackrel{(a)}{\sim} & 1-\bar{c}(f*f)(z)y^{-2/\alpha}\int_{\R^{2}}\left[1-F_{h }\left(\Vert x\Vert^{\alpha}\right)\right]\d x\end{eqnarray}
where $(a)$ follows from the Lebesgue dominated convergence theorem ($f*f$ is a very nice function  since $f$ is a PDF). So we have $\mathcal{G}\left(F_{h }\left(\frac{y}{g(.-z)}\right)\right)\sim\theta_{1}y^{-2/\alpha}$. 
For a Neyman-Scott cluster process, the second order product density
is given by \cite[p.158]{stoyan}, \[
\rho^{(2)}(r)=\lambda^{2}+\frac{\lambda\mu(r)}{\pi\bar{c}}\sum_{n=2}^{\infty}p_{n}n(n-1)\]
where $p_{n}$ is the distribution of the number of points in the
representative cluster. $\mu(r)/\pi$ denotes the density of the distribution
function for the distance between two independent random points which
were scattered using the distribution $f(x)$ of the representative
cluster. When the number of points inside each cluster is  Poisson
distributed with mean $\bar{c}$, we have $\sum_{n=2}^{\infty}p_{n}n(n-1)=\bar{c}^{2}$.
We also have $\mu(x)/\pi=(f*f)(x)$ Estimating $\varphi(y)$ we have
\begin{eqnarray}
\varphi(y) & = & \frac{1}{y\lambda}\int_{\R^{2}}g(x-z)\rho^{(2)}(x)\int_{0}^{y/g(x-z)}\nu \d F_{h }(\nu)\d x\nonumber\\
 & = & \underbrace{\frac{\lambda}{y}\int_{\R^{2}}\Vert x\Vert^{-\alpha}\int_{0}^{y\Vert x\Vert^{\alpha}}\nu \d F_{h }(\nu)d}_{T_{1}}\nonumber\\
 &  & +\underbrace{\frac{\bar{c}}{y\pi}\int_{\R^{2}}g(x-z)\mu(x)\int_{0}^{y/g(x-z)}\nu \d F_{h }(\nu)\d x}_{T_{2}}\end{eqnarray}
 By change of variables, we have \begin{eqnarray}
T_{1} & = & \frac{2\pi\lambda y^{-2/\alpha}}{\alpha-2}\int_{0}^{\infty}\nu^{2/\alpha}\d F_{h}(\nu)\end{eqnarray}
 For the term $T_{2}$,\begin{eqnarray}
T_{2} & = & \frac{\bar{c}}{y\pi}\int_{0}^{\infty}\nu \d F_{h }(\nu)\int_{\R^{2}}\Vert x-z\Vert^{-\alpha}1_{\Vert x-z\Vert^{\alpha}>vy^{-1}}\mu(x)\d x\nonumber\\
 & = & \frac{\bar{c}}{\pi y^{2/\alpha}}\int_{0}^{\infty}\nu \d F_{h}(\nu)\int_{\R^{2}}\Vert x\Vert^{-\alpha}1_{\Vert x\Vert^{\alpha}>v}   \mu\left(\frac{x}{y^{1/\alpha}}+z\right)  \d x\nonumber\\
 & \sim & \frac{\mu(z)\bar{c}}{\pi y^{2/\alpha}}\int_{0}^{\infty}\nu \d F_{h }(\nu)\int_{\R^{2}}\Vert x\Vert^{-\alpha}1_{\Vert x\Vert^{\alpha}>v}\d x\nonumber\\
 & = & \frac{y^{-2/\alpha}2\mu(z)\bar{c}}{\alpha-2	}\int_{0}^{\infty}\nu^{2/\alpha}\d F_{h}(\nu)\end{eqnarray}
 So we have $\varphi(y)\sim\theta_{2}y^{-2/\alpha}$. Hence from Theorem
\ref{thm:CCDF}, we have $\bar{F}_{I}^{l}(y)\sim\theta_{1}y^{-2/\alpha}$and
$\bar{F}_{I}^{u}(y)\sim(\theta_{1}+\theta_{2})y^{-2/\alpha}$.
\end{proof}

\subsection{Outage probability, in Poisson cluster process when the number of
cluster points are fixed.\label{sub:fixed}}

In this subsection we derive the conditional Laplace transform in
a Poisson cluster process, when the number of points in each cluster
are fixed to be $\bar{c}\in\mathbb{N}$ and $\bar{c}>0$. We also
assume that each point is independently distributed with density $f(x)$.
In this case the moment generating function of the number of points
in the representative cluster is given by\[
M(z)=z^{\bar{c}}\]
Using the same notation as in Section\foreignlanguage{english}{ }\ref{sec:Neyman-Scott-cluster-process},
and from \foreignlanguage{english}{\eqref{eq:fixed1} and }\eqref{eq:fixed2},
we have \begin{eqnarray}
E_{0}^{!}\left(\prod_{x\in\phi}v(x)\right) & = & \tilde{G}(v)\int_{N}\prod_{x\in\psi}v(x)\tilde{\Omega}_{0}^{!}(d\psi)\nonumber\\
 & = & \tilde{G}(v)\int_{N}\prod_{x\in\psi}v(x)\int_{\R^{2}}\Omega_{y}^{!}(d\psi_{y})f(y)\d y\nonumber\\
 & = & \tilde{G}(v)\int_{\R^{2}}\int_{N}\prod_{x\in\psi}v(x)\Omega_{Y}^{!}(d\psi_{y})f(y)\d y\nonumber\\
 & \stackrel{(a)}{=} & \tilde{G}(v)\int_{\R^{2}}\left(\int_{\R^{2}}v(x-y)f(x)\right)^{\bar{c}-1}f(y)\d y\end{eqnarray}
where $(a)$ follows from the fact that the points are independently
distributed and we are not counting the point at the origin. In this
case $\tilde{G}(v)$ is given by\[
\tilde{G}(v)=\exp\left\{ -\lambda_{p}\int_{\R^{2}}1-\left(\int_{\R^{2}}v(x+y)f(y)\d y\right)^{\bar{c}}\d x\right\} \]
Hence the success probability (Rayleigh fading) is given by \begin{eqnarray}
\mathbb{P}(\text{success}) & = & \exp\left\{ -\lambda_{p}\int_{\R^{2}}1-\tilde{\beta}(R,y)^{\bar{c}}\d y\right\} \int_{\R^{2}}\tilde{\beta}(R,y)^{\bar{c}-1}f(y)\d y\end{eqnarray}
where \[
\tilde{\beta}(R,y)=\int_{\R^{2}}\frac{f(x)}{1+\frac{g(R)}{T}g(x-y-z)}\d x\]

\subsection{\label{sub:nakagami}Outage probability of Nakagami-m fading}

Here, we derive the success probability when the fading distribution
  is Nakagami-m distributed. We also assume $m\in\mathbb{N}$
and $W=0$. The PDF of the  power fading coefficient $y=h_{x}$ is
given by %

\[
p(y)=\frac{1}{\Gamma(m)}\left(\frac{m}{\Omega}\right)^{m}y^{m-1}e^{-my/\Omega}\]
From (\ref{eq:main}), we have\[
\mathbb{P}(\text{success})=\mathbb{P}\left(\frac{h  g(z)}{W+I_{\phi\setminus\{0\}}(z)}\geq T\right)\]

\begin{eqnarray}
\mathbb{P}(\text{success}) & = & 1-\left(\frac{T}{g(z)}\right)^{m}\int_{0}^{\infty}\frac{1}{\Gamma(m)}\left(\frac{m}{\Omega}\right)^{m}y^{m-1}e^{-\frac{Tm}{\Omega g(z)}y}\mathbb{P}^{!0}(I_{\phi}(z)>y)\d y\end{eqnarray}
Using integration by parts we get,

\begin{eqnarray}
\mathbb{P}(\text{success}) & = & \frac{1}{\Gamma(m)}\int_{0}^{\infty}\Gamma(m,\frac{Tm}{\Omega g(z)}y)d\mathbb{P}^{!0}(I_{\phi}(z)\leq y)\nonumber \\
 & \stackrel{(a)}{=} & \frac{(m-1)!}{\Gamma(m)}\sum_{k=0}^{m-1}\frac{1}{k!}\int_{0}^{\infty}e^{-\frac{Tm}{\Omega g(z)}y}y^{k}d\mathbb{P}^{!0}(I_{\phi}(z)\leq y)\nonumber \\
 & \stackrel{(b)}{=} & \sum_{k=0}^{m-1}\frac{(-1)^{k}}{k!}\frac{d^{k}}{ds^{k}}\mathcal{L}_{I_{\phi}(z)}(s)|_{s=Tm/\Omega g(z)}\label{eq:succ-naka0}\end{eqnarray}
where $(a)$ follows from the series expansion of incomplete Gamma
function when $m$ is an integer and $(b)$ follows from the properties
of the Laplace transform and $\Gamma(m)=(m-1)!$ when $m$ is an integer.
We also have \[
\mathcal{L}_{h_{x} }(sg(x-z))=\frac{1}{(1+\frac{\Omega}{m}sg(x-z))^{m}}\]
Hence from Lemma \ref{lem:laplace-interference}, we have 

\begin{eqnarray}
  \mathcal{L}_{I_\phi(z) }(s)& = & \exp\left[-\lambda_{p}\int_{\mathbb{R}^{2}}1-\exp(-\bar{c}\bar{\beta}(s,z,y))\d y\right]\int_{\mathbb{R}^{2}}\exp(-\bar{c}\bar{\beta}(s,z,y))f(y)\d y\label{eq:success-naka}\end{eqnarray}
where \[
\bar{\beta}(s,z,y)=1-\int_{\mathbb{R}^{2}}\frac{1}{(1+\frac{\Omega}{m}sg(x-y-z))^{m}}f(x)\d x\]
For integer $m\geq1$, $\mathbb{P}(\text{success})$ can be evaluated
from (\ref{eq:succ-naka0}) and (\ref{eq:success-naka}). For $m=1$,
the probability evaluated from (\ref{eq:succ-naka0}) and (\ref{eq:success-naka})
matches that of Lemma \ref{lem:outage0}. %
{}%
{}

\subsection{\label{sub:Proof-of-Lemma-limit} Proof of Lemma \ref{lem:limit}}

\begin{proof}
\begin{eqnarray}
\frac{P_{p}(\lambda_{p}\bar{c})}{P_{1}(R,\bar{c},\lambda_{p})} & = & \exp\Big[-\lambda_{p}\bar{c}\int_{\R^{2}}\beta(R,y)\d y+\lambda_{p}\int_{\R^{2}}(1-\exp[-\bar{c}\beta(R,y)])\d y\Big]\nonumber\\
 & = & \exp\Big[-\lambda_{p}\int_{\R^{2}}\{\underbrace{\bar{c}\beta(R,y)-1+\exp\Big[-\bar{c}\beta(R,y)\Big]}_{\nu(R,y)}\}\d y\Big]\end{eqnarray}
 Since $1-\exp(-ax)\leq ax$, we have that $\nu(R,y)>0$. We also
have from the dominated convergence theorem and \eqref{eq:assume2}\[
\lim_{R\rightarrow\infty}\nu(R,y)=\frac{\bar{c}}{1+T^{-1}}-1+\exp\left(-\frac{\bar{c}}{1+T^{-1}}\right)>0\]
 which is a constant. So using Fatou's lemma~\cite{folland} ($\liminf\int f_{n}\geq\int(\liminf f_{n}),~f_{n}>0$),
we have \begin{eqnarray}
\lim_{R\rightarrow\infty}\frac{P_{p}(\lambda_{p}\bar{c})}{P_{1}(R,\bar{c},\lambda_{p})} & = & \exp[-\lambda_{p}\lim_{R\rightarrow\infty}\int_{\R^{2}}\nu(R,y)\d y]\nonumber\\
 & \leq & \exp[-\lambda_{p}\int_{\R^{2}}\lim_{R\rightarrow\infty}\nu(R,y)\d y]\nonumber\\
 & = & \exp[-\lambda_{p}\infty]=0\end{eqnarray}

\end{proof}

\subsection{\label{sub:Proof-of-Lemma4} Proof of Lemma \ref{lemma4}}

\begin{proof}
From \eqref{eq:onnne}, the probability of success is \begin{eqnarray}
\mathbb{P}(\text{success}) & = & P_{p}(\lambda_{p}\bar{c})\underbrace{\exp\Big[\lambda_{p}\bar{c}\int_{\R^{2}}\beta(R,y)\eta(\bar{c},R,y)\d y\Big]}_{T_{1}}\underbrace{P_{2}(R,\bar{c})}_{T_{2}}\end{eqnarray}
 where $\eta(\bar{c},R,y)=\nu(\bar{c}\beta(R,y))$ and $\nu(x)=(\exp(-x)-1+x)/x$
an increasing function of $x$. From Young's inequality ~\cite[Sec. 8.7]{folland}
we have $\beta(R,y)\leq\min\{1,\sup\{f(x)\}R^{2}T^{2/\alpha}C(\alpha)\}$.
Hence\[
\eta(\bar{c},R,y)\leq\nu(\bar{c}\min\{1,\sup\{f(x)\}R^{2}T^{2/\alpha}C(\alpha)\})\]
 With a slight abuse of notation, let $\eta(\bar{c},R)=\nu(\bar{c}\min\{\sup\{f(x)\}R^{2}T^{2/\alpha}C(\alpha),1\})$.
Hence \begin{eqnarray}
T_{1} & \leq & \exp[\lambda_{p}\bar{c}\int_{\R^{2}}\beta(R,y)\eta(\bar{c},R)\d y]\nonumber \\
 & = & \exp[\lambda_{p}\bar{c}T^{2/\alpha}R^{2}\eta(\bar{c},R)C(\alpha)]\label{upper1}\end{eqnarray}
 Also observe that $\eta(\bar{c},R)\lessapprox R^{2}$. So $T_{1}\lessapprox\exp(R^{4})$.
\begin{eqnarray}
T_{2} & = & \int_{\R^{2}}1-\bar{c}\beta(R,y)+\bar{c}\beta(R,y)\sum_{k=2}^{\infty}\frac{(-1)^{n}}{n!}(\bar{c}\beta(R,y))^{n-1}f(y)\d y\nonumber\\
 & \leq & \int_{\R^{2}}[1-\bar{c}\beta(R,y)+\bar{c}\beta(R,y)\eta(\bar{c},R)]f(y)\d y\nonumber\\
 & = & 1-[1-\eta(\bar{c},R)]\bar{c}\int_{\R^{2}}\beta(R,y)f(y)\d y\end{eqnarray}
 If one considers $x$ and $y$ as identical and independent random
variables with density functions $f$, we then have $\int_{\R^{2}}\beta(R,y)f(y)\d y=E[\frac{1}{1+\frac{g(R)}{T}\Vert x-y-R\Vert^{\alpha}}]$.
Let $0<\kappa<1$ be some constant. Using the Chebyshev inequality
we get \begin{eqnarray}
E\left[\frac{1}{1+\frac{g(R)}{T}\Vert x-y-R\Vert^{\alpha}}\right] & \geq & \kappa P\left[\frac{1}{1+\frac{g(R)}{T}\Vert x-y-R\Vert^{\alpha}}\geq\kappa\right]\nonumber\\
 & = & \kappa P\Big[\Vert x-y-R\Vert\leq(\frac{1}{\kappa}-1)^{1/\alpha}RT^{1/\alpha}\Big]~~(**)\end{eqnarray}
 The PDF of $z=x-y$ is given by $(f*f)(z)$, since $y$ is rotation-invariant. 
 Choosing $\kappa=T/(1+T)$ we have \begin{eqnarray}
(**) & = & \frac{T}{1+T}\int_{B(R,R)}(f*f)(x)\d x\nonumber\\
 & \geq & \frac{T}{1+T}\int_{B(R,R)}\Vert x\Vert \d x\nonumber\\
 & = & R^{3}\underbrace{\frac{T}{1+T}\int_{B(1,1)}\Vert x\Vert \d x}_{C_{2}}\end{eqnarray}
 So we have \begin{eqnarray}
P_{2} & \leq & 1-[1-\eta(\bar{c},R)]R^{3}C_{2\nonumber}\\
 & \lessapprox & 1-R^{3}+R^{5}\end{eqnarray}
 Also we have $T_{1}\lessapprox\exp(R^{4})\lessapprox1+1.01R^{4}$.
So we have $P_{2}T_{1}\lessapprox1-R^{3}+R^{5}-1.01R^{7}+1.01R^{9}<1$
for small $R\neq0$. {\em Hence for small $R$ we have $\mathbb{P}(\text{success})\leq P_{p}(\lambda_{p}\bar{c})$.}\\
 
\end{proof}

\subsection{\label{sub:Proof-upper-cap}Proof of Lemma \ref{lem:upper-cap}}

\begin{proof}
We find $C_{u}(\epsilon,T)$ and hence upper bound the transmission
capacity. We have from the derivation of Lemma~\ref{lemma4} \begin{eqnarray}
P(\lambda,T) & \leq & P_{p}(\lambda_{p}\bar{c})\exp[\lambda_{p}\bar{c}\beta_{I}\eta(\bar{c},R)]P_{2}(R,\bar{c})=P_{u}(\bar{c}\lambda_{p},T)\end{eqnarray}

where $\eta(\bar{c},R)=(\exp(-\bar{c}\hat{\beta})-1+\bar{c}\hat{\beta})/\bar{c}\hat{\beta}$.
With $A_{u}=\{\lambda_{p}\bar{c},\: P_{u}(\lambda_{p}\bar{c},T)\geq1-\epsilon\}$,
it is sufficient to prove $\sup A_{u}\leq C_{p}(\epsilon,T)$.\foreignlanguage{english}{
Also observe that} $P_{u}(\bar{c}\lambda_{p},T)\rightarrow0$\foreignlanguage{english}{
as }$\bar{c}\rightarrow\infty$\foreignlanguage{english}{ independent
of }$\lambda_{p}$\foreignlanguage{english}{. Hence we can assume
}$\bar{c}$\foreignlanguage{english}{ is finite for the proof.} We
proceed by contradiction.

Let $\sup A_{u}>C_{p}(\epsilon,T)$. \foreignlanguage{english}{Hence
there exists a $\delta>0$, $\lambda_{p}\geq0,\ \bar{c}\geq0$ such
that $\lambda_{p}\bar{c}=(C_{p}(\epsilon,T)/(1-\epsilon))+\delta\in A_{u}$.
At this value of $\lambda_{p}\bar{c}$ we have \begin{eqnarray}
1-\epsilon\leq P_{u}(\bar{c}\lambda_{p},T) & = & (1-\epsilon)P_{p}(\delta)\exp[\eta(\bar{c},R)\{\ln(\frac{1}{1-\epsilon})+\delta\beta_{I}\}]P_{2}(R,\bar{c})\nonumber \\
 & = & (1-\epsilon)^{1-\eta(\bar{c},R)}\underbrace{P_{p}(\delta(1-\eta(\bar{c},R)))}_{T_{1}}P_{2}(R,\bar{c})\label{eq:121}\end{eqnarray}
 From the derivation of Lemma~\ref{lemma4}, we have $P_{2}(R,\bar{c})\leq1-[1-\eta(\bar{c},R)]\bar{c}k$,
with equality only when $\bar{c}=0.$ Hence we have \begin{eqnarray}
p_{u}(\bar{c}\lambda_{p},T) & \leq & T_{1}(1-\epsilon)^{1-\eta(\bar{c},R)}(1-[1-\eta(\bar{c},R)]\bar{c}k)\end{eqnarray}
 Since $\frac{\exp(-x)-(1-x)}{x}\leq\frac{x}{1+x}$, we have $\eta(\bar{c},R)\leq\bar{c}\hat{\beta}/(1+\bar{c}\hat{\beta})$.
Using the upper bound for $\eta(\bar{c},R)$ \begin{eqnarray}
p_{u}(\bar{c}\lambda_{p},T) & \leq & T_{1}(1-\epsilon)(1-\epsilon)^{-\frac{\bar{c}\hat{\beta}}{1+\bar{c}\hat{\beta}}}\left(1-\left[1-\frac{\bar{c}\hat{\beta}}{1+\bar{c}\hat{\beta}}\right]\bar{c}\hat{\beta}\rho(T)\right)\nonumber\\
 & = & T_{1}(1-\epsilon)\underbrace{(1-\epsilon)^{-\frac{\bar{c}\hat{\beta}}{1+\bar{c}\hat{\beta}}}\left(1-\frac{\bar{c}\hat{\beta}\rho(T)}{1+\bar{c}\hat{\beta}}\right)}_{T_{2}}\end{eqnarray}
 Using the inequality $1-ay\leq(1-b)^{y},~b\leq1-e^{-a},y\geq0$,
substituting $y=\frac{\bar{c}\hat{\beta}}{1+\bar{c}\hat{\beta}},~b=\epsilon,a=\rho(T)$,
we get $T_{2}\leq1$. Hence we have \begin{eqnarray}
p_{u}(\bar{c}\lambda_{p},T) & \leq & (1-\epsilon)P_{p}(\delta(1-\eta(\bar{c},R)))\label{eq:122}\end{eqnarray}
 So if $\delta>0$, and $\bar{c}$ finite}, we also have\foreignlanguage{english}{
$P_{p}(\delta(1-\eta(\bar{c},R)))<1$. So we have a contradiction
from (\ref{eq:121}) and (\ref{eq:122}).} Hence there exists no such
$\delta$ and hence $\sup A_{u}\leq C_{p}(\epsilon,T)$. \foreignlanguage{english}{We
can achieve $C_{u}(\epsilon,T)=C_{p}(\epsilon,T)$, by using $\lambda_{p}=n\frac{C_{p}(\epsilon,T)}{1-\epsilon}-1,\bar{c}=1/n$
for $n$ very large. As $n\rightarrow\infty,P_{u}(\bar{c}\lambda_{p},T)\rightarrow P_{p}(\bar{c}\lambda_{p},T)$
.}
\end{proof}

\end{appendix}
\end{document}